%% file: main.tex
\documentclass[conference]{IEEEtran}

\ifCLASSINFOpdf
\else
\fi
\usepackage{algorithm}
\usepackage{algpseudocode}
\usepackage{hyperref}
\usepackage{caption}
\usepackage{subcaption}
\usepackage{amsmath}
\usepackage{graphicx}
\usepackage{soul}

\usepackage{mdframed}
\usepackage{enumitem}
\usepackage{multirow}
\usepackage{xcolor}
\usepackage[normalem]{ulem}
\usepackage{balance}
\usepackage{amssymb}

\usepackage[inkscapelatex=false]{svg}
\usepackage{array}

\begin{document}
%
\title{\textit{Know Me by My Pulse}: Toward Practical Continuous Authentication on Wearable Devices via Wrist-Worn PPG}

\author{
    Wei Shao$^{1}$, Zequan Liang$^{1}$, Ruoyu Zhang$^{1}$, Ruijie Fang$^{1}$, Ning Miao$^{1}$, Ehsan Kourkchi$^{1}$, Setareh Rafatirad$^{1}$, \\
    Houman Homayoun$^{1}$, and Chongzhou Fang$^{2,*}$\thanks{$^*$This work was primarily conducted while the author was affiliated with University of California, Davis.}\\%
    Email: \{wayshao, zqliang, ryuzhang, rjfang, nmiao, ekay, srafatirad, hhomayoun\}@ucdavis.edu, and\\cxfeec@rit.edu\\%
    $^{1}$University of California, Davis\\ %
    $^{2}$Rochester Institute of Technology\\%
}
	

%


\IEEEoverridecommandlockouts
\makeatletter\def\@IEEEpubidpullup{6.5\baselineskip}\makeatother
\IEEEpubid{\parbox{\columnwidth}{
This is the camera-ready version accepted at NDSS 2026.\\
The official publication will be available at: https://www.ndss-symposium.org
}
\hspace{\columnsep}\makebox[\columnwidth]{}}

\maketitle
\pagestyle{plain}
\begin{abstract}
Biometric authentication using physiological signals offers a promising path toward secure and user-friendly access control in wearable devices. While electrocardiogram (ECG) signals have shown high discriminability, their intrusive sensing requirements and discontinuous acquisition limit practicality. Photoplethysmography (PPG), on the other hand, enables continuous, non-intrusive authentication with seamless integration into wrist-worn wearable devices. However, most prior work relies on high-frequency PPG (e.g., 75--500\,Hz) and complex deep models, which incur significant energy and computational overhead—impeding deployment in power-constrained real-world systems.
In this paper, we present the first real-world implementation and evaluation of a continuous authentication system on a smartwatch, We-Be Band, using low-frequency (25\,Hz) multi-channel PPG signals. Our method employs a Bi-LSTM with attention mechanism to extract identity-specific features from short (4\,s) windows of 4-channel PPG. Through extensive evaluations on both public datasets (PTTPPG) and our We-Be Dataset (26 subjects), we demonstrate strong classification performance with an average test accuracy of 88.11\%, macro F1-score of 0.88, False Acceptance Rate (FAR) of 0.48\%, False Rejection Rate (FRR) of 11.77\%, and Equal Error Rate (EER) of 2.76\%. Our 25\,Hz system reduces sensor power consumption by 53\% compared to 512\,Hz and 19\% compared to 128\,Hz setups without compromising performance. We find that sampling at 25\,Hz preserves authentication accuracy, whereas performance drops sharply at 20\,Hz while offering only trivial additional power savings, underscoring 25\,Hz as the practical lower bound. Additionally, we find that models trained exclusively on resting data fail under motion, while activity-diverse training improves robustness across physiological states.
\end{abstract}


%
\IEEEpeerreviewmaketitle

\section{Introduction}
Authentication is a critical part of modern digital systems, ensuring that only authorized users can access sensitive information and resources \cite{phadke2013importance,brandao2018importance,simmons1988survey}. Traditional authentication methods, e.g. passwords, and PINs, have been proven to be vulnerable to breaches and often inconvenient \cite{garrett2015vulnerability,chanda2016password}. In the world of wearable IoT devices, where users' sensitive real-time data are being continuously collected and processed on the device, it is important to provide seamless authentication protection mechanisms beyond traditional authentication approaches. 

Biometric authentication is a compelling approach to address this challenge, relying on the distinct physiological or behavioral patterns of different individuals, which can be considered reliable sources of authentication \cite{weaver2006biometric,bhattacharyya2009biometric,rui2018survey}. Cardiac signal-based approaches, e.g. electrocardiogram (ECG)-based and photoplethysmography (PPG)-based authentication, are especially attractive because they can be integrated into real-time systems to enable continuous authentication~\cite{uwaechia2021comprehensive,sancho2018biometric,zhao2021robust}. ECG-based biometrics have shown high accuracy in research; however, ECG signal collection is often hampered by its intrusive nature and discontinuous acquisition (typically requiring contact electrodes or active user engagement, such as placing a finger on a sensor).


\begin{figure}
    \centering
    \includegraphics[width=\linewidth]{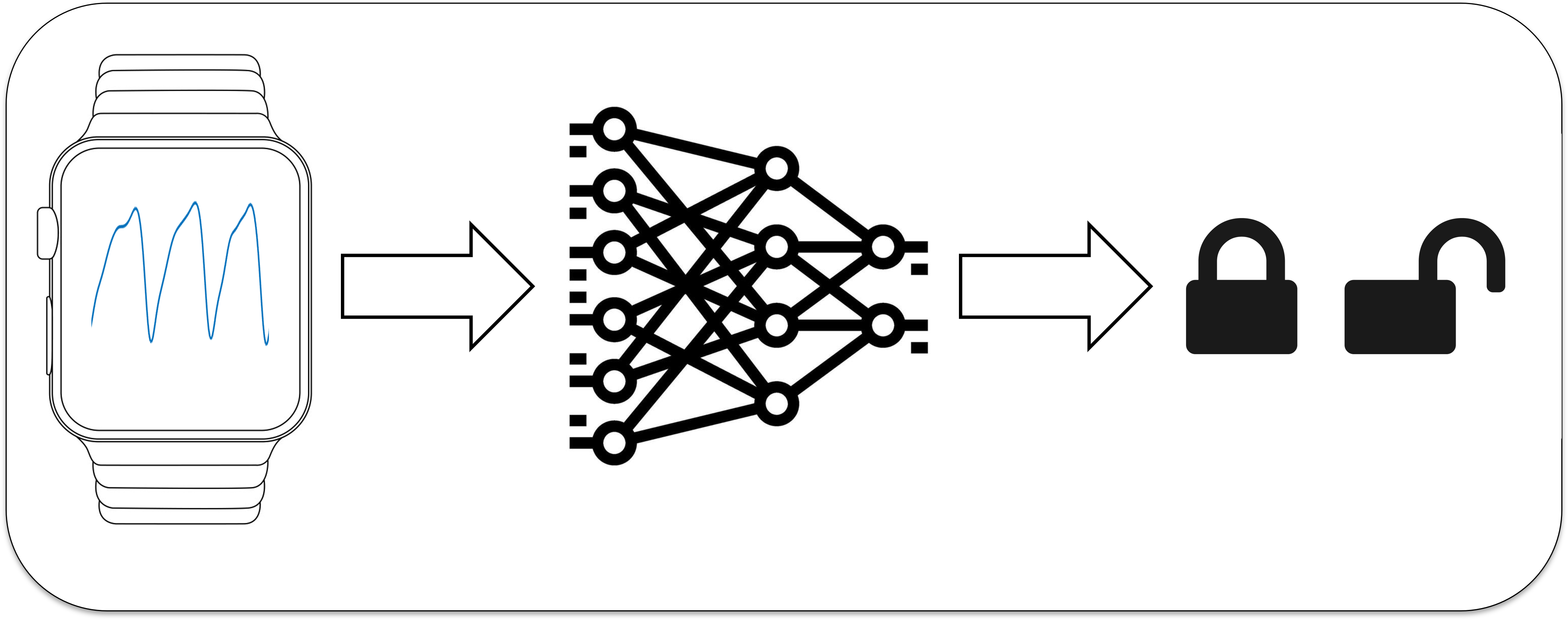}
    \caption{Authentication System Overview}
    \label{systemOverview}
\end{figure}

PPG, by contrast, is easily captured by the optical sensors already embedded in most wearables, making it non-intrusive, ubiquitous, and power-efficient. Unlike ECG, which demands low‑impedance skin–electrode contact and thus imposes more hardware complexity and user burden, PPG can be sensed passively with minimal user involvement. Although ECG yields cleaner waveforms and zero phase lag because it measures the heart’s electrical impulse directly, these benefits come at the cost of comfort and form‑factor flexibility. PPG’s main drawbacks are a higher susceptibility to motion artifacts, ambient‑light interference, and a physiological lag behind the cardiac electrical event, all of which can degrade signal quality in everyday use. Nevertheless, early studies that used high‑frequency PPG have reported excellent authentication accuracy under controlled conditions~\cite{zhang2023secure,7791193,8411233}.

Despite PPG’s potential, gaps remain in the current literature on continuous PPG-based authentication as well as in a practical way. These solutions frequently fall short in enabling continuous, real-world operation due to practical constraints like power consumption and comfort, and often rely on high-quality PPG signals collected at high sampling rates (typically 75--500\,Hz) using professional or medical-grade sensors~\cite{8350983,8607019,8553585}. Such high-fidelity data capture rich heartbeat morphology but at the cost of increased power consumption and data volume. This is impractical for power-constraint wearable devices. Moreover, much of the past work has been evaluated on pre-collected datasets or in laboratory scenarios with stationary subjects. Practical real-world validation of PPG authentication has been limited. Few studies attempted on-device deployment of PPG authentication algorithms, and they often do not address power and energy impacts.

In this paper, we address these challenges by proposing a continuous authentication system using low-frequency multi-channel PPG signals and evaluate the system through practical real-world demonstration, which to the best of our knowledge, is the first work to fully implement and evaluate a continuous biometric authentication system on a smartwatch. Figure~\ref{systemOverview} shows the high-level overview of the authentication system. We chose to rely solely on the PPG signal. While incorporating additional sensors like Galvanic Skin Response (GSR), accelerometer, or even ECG could help, our aim is to demonstrate strong results using just PPG. Our approach leverages the We-Be Band---a research-oriented smartwatch equipped with PPG and other physiological sensors~\cite{WeBeIntro,we-be-band}---to collect photoplethysmogram data from users. Uniquely, we sample the PPG at native 25\,Hz, substantially lower than in most prior studies, to enable ultra-low-power operation. By using multi-channel PPG (multiple wavelengths/sensors), we capture complementary cardiovascular features even at this lower sampling rate. These signals are fed into a machine learning model that learns each user’s heartbeat pattern. The model is lightweight and is suitable for real-time inference on the edge device or wearable’s embedded processor. Our system thus emphasizes an energy-aware design: it maintains robust authentication accuracy while minimizing sampling frequency and computation, thereby optimizing for battery life of wearable devices.

To evaluate our approach, we aim to answer the following key Research Questions (RQs):

\begin{itemize}
\label{RQs}
    \item \textbf{RQ1:} How does PPG sampling frequency affect authentication accuracy and generalization in real-world wearable use cases?
    \item \textbf{RQ2:} To what extent does multi-channel PPG improve authentication resilience against impersonation and signal noise compared to single-channel input?
    \item \textbf{RQ3:} What is the impact of activity-aware training (e.g., walking, typing) on the robustness of PPG-based biometric models in dynamic environments?
    \item \textbf{RQ4:} Can a low-power authentication model operating at 25\,Hz PPG sampling achieve accuracy and robustness comparable to high-frequency counterparts on real-world wearable devices?
    \item \textbf{RQ5:} How feasible is real-time, continuous biometric authentication using low-frequency PPG on embedded devices in terms of power and usability?
\end{itemize}

We answer these questions through extensive experimentation with our prototype system. In terms of authentication performance, our results show that the proposed 25\,Hz multi-channel PPG approach can achieve strong accuracy---for instance, about 88\% classification accuracy in identifying users, with an equal error rate (EER) below 3\%. This is comparable to state-of-the-art methods that use significantly higher sampling rates, indicating that a higher rate is not strictly necessary for effective biometric recognition. Importantly, the system maintains this accuracy in realistic conditions. We validate it with users during various activities and over multiple sessions to demonstrate its ability for real-world use. The trained deep learning model runs in real-time, and we successfully deployed the pipeline on a prototype with the We-Be Band for continuous operation. Equally significant are the implications for energy efficiency. By using a 25\,Hz sampling rate instead of the common 75+\,Hz, our system drastically reduces the volume of data and frequency of sensor readings, directly lowering the power consumption of both the PPG sensor and the processing workload. We estimate that this down-sampling approach leads to considerable power savings. In our measurements, the power consumption for PPG sensing alone at 25\,Hz is 53\% less than at 512\,Hz and 19\% less than at 128\,Hz. Combined with the power saved by processing lower-frequency data this leads to a longer battery life consistent with multi-day wearable use.

Additionally, we gained two notable insights from our experiments. First, we show that a lower sampling rate can sometimes yield similar accuracy before it reaches 20\,Hz. Lower rates starting at 20 Hz quickly compromise the performance, outweighing any additional energy benefit. In certain scenarios, the model’s generalization ability is similar on down-sampled PPG as the higher rates, likely because the lower rate filters out high-frequency noise and forces the model to focus on the most salient features of the pulse waveform. Further discussion can be found in Section~\ref{sec:discussion-sampling}. Second, incorporating diverse activities in the training data (rather than only resting-state PPG) greatly improves the system’s robustness and biometric separability. Models trained on PPG data across a range of activities (e.g., walking, typing, exercising) were significantly better at distinguishing users under various real-life conditions, highlighting that activity-induced variations can enrich the biometric feature set (we discuss this further in Section~\ref{sec:activity-diversity-dis}).

In summary, the contributions of this paper are as follows:
\begin{itemize}
  \item \textbf{A low-frequency, multi-channel PPG authentication system:} We design and implement a continuous authentication pipeline based on 25\,Hz multi-channel PPG signals, demonstrating that robust performance can be achieved without high-frequency sensing.
  
  \item \textbf{Empirical validation of sampling rate impact:} Through downsampling experiments on public datasets, we show that lower sampling rates can retain or even improve authentication accuracy while dramatically reducing power consumption.
  
  \item \textbf{Practical Real-world implementation and evaluation:} Our model achieved 88.11\% accuracy, 0.48\% False Acceptance Rate (FAR), 11.77\% False Rejection Rate (FRR), and Equal Error Rate (ERR), 2.76\%. We further design and deploy a prototype system that supports the entire pipeline of authentication tasks in real-world scenarios involving human participants on a smartwatch (We-Be Band). 
  
  \item \textbf{Insights into activity-aware training:} We demonstrate that training the model on a diverse set of activities (e.g., sitting, walking, typing) significantly improves generalization across conditions, countering the long-standing bias toward resting-state biometrics.
  
  \item \textbf{Design guidance for low-power wearable devices:} We provide power measurements and performance benchmarks to guide developers in selecting sampling rates that balance biometric fidelity and energy efficiency for continuous authentication applications.
\end{itemize}

\section{Background}

Biometric authentication has evolved from relying on traditional physical traits, such as fingerprints and facial features, to utilizing physiological signals for enhanced security and robustness. Among these, electrocardiogram (ECG) and photoplethysmography (PPG) have garnered significant attention due to their uniqueness and feasibility for continuous monitoring. This section provides foundational context for our work, including key challenges and motivations behind our multi-channel, low-power PPG-based approach. For an expanded discussion of related literature, please refer to Section~\ref{sec:related_work}.

\subsection{ECG-Based Authentication}
ECG signals have been widely studied for biometric authentication due to their unique cardiac waveform characteristics~\cite{ECGAMD,ECGA,ECGA_ANA,ECGA_MWD,ECG_Survey}. However, ECG acquisition typically requires electrodes and direct skin contact, making it less suitable for continuous, unobtrusive use in daily life. Furthermore, ECG signals are not continuously captured in most wearable devices, limiting their practicality for real-time authentication systems.

\subsection{Photoplethysmography (PPG) as an Alternative}
Photoplethysmography (PPG) offers a non-invasive and wearable-friendly alternative to ECG. Early works using high-frequency PPG (75--500\,Hz) demonstrated excellent authentication performance in controlled environments \cite{zhang2023secure,7791193,8411233}. However, such systems often incur substantial power and computational overhead, which can hinder scalability to continuous real-world use. These limitations have prompted research into lower-frequency, multi-channel, and hardware-aware approaches to preserve both accuracy and efficiency. Table~\ref{tab:ppg_auth_lit} summarizes prior works on PPG-based biometric authentication.

\begin{table*}[ht!]
  \centering
  \caption{Representative Prior Works on PPG-Based Biometric Authentication}
  \label{tab:ppg_auth_lit}
  {\fontsize{10}{12}\selectfont
  \begin{tabular}{
  >{\centering\arraybackslash}m{1cm}  
  >{\centering\arraybackslash}m{2cm}  
  >{\centering\arraybackslash}m{2.5cm}  
  >{\centering\arraybackslash}m{2cm}  
  >{\raggedright\arraybackslash}m{8.5cm}  
}
    \hline
    \textbf{Study} & \textbf{Model Type} & \textbf{Sampling Rate} & \textbf{Dataset} & \textbf{Key Contribution} \\
    \hline
    \cite{8411233} & CWT + DLDA & 100--300\,Hz & BioSec.Lab PPG & Evaluated performance across states (e.g., exercise, stress). \\
    \hline
    \cite{6004938} & Statistical Analysis & 75\,Hz & Custom (finger sensor) & Demonstrated feasibility of PPG for ID under controlled, low-motion conditions. \\
    \hline
    \cite{8350983} & CNN + LSTM & 125\,Hz & TROIKA & Achieving 96\% accuracy on wrist-worn PPG. \\
    \hline
    \cite{8553585} & CNN & 125\,Hz & TROIKA, PulseID & End-to-end PPG authentication with AUC of 78--83\%. \\
    \hline
    \cite{zhao2021robust} & Random Forest + Wavelet & 300\,Hz & Custom & Continuous auth with cardiac features from wrist-worn wearable devices. \\
    \hline
    \cite{10109506} & Random Forest & 125\,Hz & Custom & Gesture-aware 2FA using PPG; proposed multimodal fusion on wearable devices. \\
    \hline
    \cite{9413906} & CNN (Fusion) & 100--300\,Hz & PRRB, TROIKA & Multi-channel fusion for variation-stable biometric recognition. \\
    \hline
    \cite{4353358} & Statistical analysis & 300\,Hz & Custom (3 subjects) & Demonstrated that derivatives of PPG signals can precisely describe individual features for identification purposes. \\
    \hline
    \cite{6656145} & Correlation-based & 75\,Hz & Custom (44/14 subjects) & Proposed continuous authentication using PPG; achieved EERs of 5.29\% and 13.47\% over different durations. \\
    \hline
    \cite{zhang2023secure} & Random Forest & 500\,Hz & PRRB, MIMIC-III, CapnoBase & Achieved 96.4\% accuracy and EER of 2.14\%. \\
    \hline
    \cite{jindal2016adaptive} & Deep Belief Network & 125\,Hz & TROIKA & Applied adaptive deep learning for PPG-based identification. \\
    \hline
    \cite{7791193} & LDA, QDA, Mahalanobis & 512\,Hz & DEAP & Implemented biometric authentication using PPG signals; achieved 90\% and 95\% accuracy for 2s and 8s signals, respectively. \\
    \hline
    \cite{zhao2020trueheart} & Gradient Boosting Tree & 300\,Hz & Custom (20 subjects) & Developed TrueHeart, a continuous authentication system using wrist-worn PPG; achieved 90\% accuracy and 4\% FDR under motion. \\
    \hline
    \cite{hwang2020evaluation} & CNN + LSTM & 100\,Hz, 300\,Hz & BioSec1, BioSec2, PRRB & Evaluated time-stable PPG biometrics across sessions; achieved 98\% accuracy in single-session and 87.1\% in two-session setups. \\
    \hline
    \cite{biswas2019cornet} & CNN + LSTM & 125\,Hz & TROIKA, IMEC-Db & Introduced CorNET for ambulant biometric ID and HR estimation; achieved 96\% ID accuracy and 1.47$\pm$3.37 BPM HR error. \\
    \hline
  \end{tabular}
  }
\end{table*}

\subsection{Sensor Hardware and PPG Channel Characteristics}
\label{sec:sensor-channels}
Our system is based on the SFH-7072 optical sensor equipped on We-Be Band, which features four PPG channels: two green (526\,nm), one red (660\,nm), and one infrared (950\,nm). Each wavelength penetrates to a different tissue depth and interacts with distinct physiological layers, enabling richer signal acquisition and improving resilience to motion artifacts and skin tone variability. This configuration is particularly advantageous for authentication tasks, which require reliable signal diversity.

\section{Methodology}
In this section, we present our continuous authentication methodology, encompassing data acquisition, preprocessing, model design, training, and evaluation.

Our approach features four key innovations: (1) a low-power PPG sensing pipeline with a 25\,Hz sampling rate, (2) the use of multi-channel PPG inputs for enhanced signal richness, (3) a class-weighted loss function to address imbalanced training data, and (4) a rigorous comparison and selection of two deep learning architectures (a Bi-LSTM with attention vs. a Transformer) to identify the optimal model for this task.

\subsection{Wearable Platform: We-Be Band}
\label{sec:webe-band}

Our data were collected using the We-Be Band, a smartwatch platform developed for low-power, multi-modal physiological monitoring. Sensors relevant to our system include:

\begin{itemize}
  \item \textbf{multi-channel PPG Sensor:} Four optical channels (two green, one red, one infrared) sampled at 25\,Hz.
  \item \textbf{Data Acquisition:} An nRF52840 SoC with ARM Cortex-M4 manages synchronized sampling and stores data on a 256\,MB flash module. BLE is supported for optional real-time streaming.
\end{itemize}

\subsection{Datasets}
\label{sec:datasets}

We evaluated our system using two datasets:  one public dataset, Pulse Transit Time PPG Dataset (PTTPPG), and one custom dataset collected using the We-Be Band.

\subsubsection{Pulse Transit Time PPG Dataset (PTTPPG)}
\label{sec:pttppg}

This public dataset~\cite{PTTPPG} includes synchronized ECG and multi-channel PPG data from 22 healthy subjects during sitting, walking, and running sessions. ECG was recorded at 500\,Hz, and PPG at 500\,Hz using six MAX30101-based channels (pleth\_1–pleth\_6). All signals are temporally aligned to within 2\,ms. We use both ECG and PPG data from this dataset for baseline comparisons across activity states.

\subsubsection{Custom Dataset: We-Be Dataset}
\label{sec:webe-dataset}

We use a multi-modal dataset from 26 volunteers (mean age: $23 \pm 3.1$ years; 38\% female) collected using the We-Be Band. Subjects wore the band on the non-dominant wrist and performed daily tasks including rest, walking, talking, etc. PPG was sampled at 25\,Hz from all four channels. This design captures a range of physiological responses and motion-induced variations. The We-Be Band’s extended battery life and onboard storage enabled uninterrupted recording across all sessions.

\subsection{PPG Signal Selection}
We select multiple PPG channels to ensure spatial and spectral diversity across skin depths, enabling the model to learn more robust and individualized signal representations. multi-channel fusion enables the extraction of complementary biometric features—such as amplitude ratios, peak timings, and pulse shape—which may not be consistently present in single-channel signals. While some studies have explored additional sensor modalities such as skin temperature and galvanic skin response (GSR), we focus on optimizing the use of multi-channel PPG for practical, and continuous authentication. 

To evaluate the impact of spectral diversity on biometric modeling and validate our selection of multi-channel configuration, we conducted an ablation study using the We-Be Dataset. Our model was first trained and evaluated in: (i) a \textbf{single-channel} setting using only one green PPG signal, and (ii) a \textbf{multi-channel} setting incorporating all four available channels.

Green light was chosen for the single-channel baseline as it is the most commonly used wavelength in commercial wearable devices, offering strong superficial pulse signals and high signal-to-noise ratios. However, this wavelength probes only shallow tissue layers and may not capture deeper or complementary hemodynamic features.

\subsection{Data Preprocessing}
The raw PPG signals are first processed with a second-order band-pass filter (0.5--12\,Hz) to remove motion artifacts, baseline drift, and high-frequency noise. Each filtered signal is then segmented into 4-second windows with 50\% overlap, producing fixed-length inputs that preserve temporal continuity while improving data density. Within each segment, the signals are Z-score normalized to eliminate amplitude variance across users and sessions, facilitating convergence during training.

\subsection{Design Choices}
\subsubsection{Window Length}\label{sec:win-overlap} 
For window length, we use 4-second windows. We evaluated 2-second, 4-second, and 8-second windows on the We-Be dataset while keeping every other component—preprocessing, network architecture, and training schedule—unchanged. The 4-second window delivered the highest accuracy (88.11\%) over the alternatives), whereas 2-second and 8-second reached 86.45\% and 86.02\%, respectively. Short 2-second windows often contain only one or two heartbeats; the reduced morphological context limits discriminative cues and increases variance. Conversely, 8-second segments bundle 5–15 beats, which inflates input dimensionality, slows training, and introduces greater intra-window variability, all of which manifested as pronounced oscillations in the learning curves. A 4-second window therefore strikes the best balance: it spans several cardiac cycles across a wide heart-rate range (35–180 bpm), yet keeps authentication latency under 5 seconds—acceptable for smartwatch unlock scenarios.

\subsubsection{Overlap}
For overlap between windows, we employ a 50\% overlap (stride = 2 s) for both training and inference. This choice is guided by three practical considerations: 
\begin{itemize}
    \item \textbf{Boundary completeness}: Any beat that straddles a window edge is fully captured in at least one neighbouring segment, mitigating clipping artifacts without resorting to beat-aligned segmentation.
    \item \textbf{Sample efficiency vs. redundancy}: Halving the stride doubles the effective dataset size and diversifies mini-batches, yet keeps adjacent windows decorrelated enough for the optimizer to see non-redundant gradients. Heavier overlap (75\%) would magnify storage and compute costs with diminishing information gain; lighter overlap (25\%) risks undersampling transient morphologies. 
    \item \textbf{Real-time update rate}: During deployment the classifier emits a decision every 2s, providing a smooth, promptly updating confidence stream suitable for majority-vote fusion and quick lock/unlock feedback. Together, the 4s window with 50\% overlap yields a responsive, computationally tractable pipeline while preserving the physiological detail necessary for reliable PPG-based authentication.
\end{itemize}

\subsubsection{Sampling Rate}
We adopt a 25\,Hz sampling rate for all baseline experiments. The impact of alternative rates is examined in Section~\ref{sec:rq1-sampling-rate} as part of RQ1.

\subsection{Deep Learning Model Architectures}
We evaluated two deep learning architectures for user classification from multi-channel PPG: a bidirectional LSTM model with attention, and a Transformer-based encoder. We choose these two models because both of them capture temporal dependencies in the sequential physiological signals, with LSTM excelling at modeling local patterns efficiently and Transformers leveraging global attention for more comprehensive feature learning. We present our comparison results and final decision later on.

Each 4-second segment contains 100 time steps (at 25\,Hz) and 4 channels (two green, one red, one infrared). The models output a probability distribution over $M=26$ users.

\subsubsection{BiLSTM+Attention model} This architecture employs stacked bidirectional LSTM layers to learn forward and backward temporal dynamics in the PPG sequence. Each time step’s hidden state is passed to a global attention layer, which computes a context vector as a weighted sum of all hidden states. The attention mechanism enables the model to focus on salient waveform segments relevant to individual identity, such as unique pulse morphology. The context vector is fed into a fully connected layer with softmax activation for user classification. The model is lightweight and suitable for on-device inference.

\subsubsection{Transformer model} The Transformer replaces recurrence with multi-head self-attention. After embedding the input and adding positional encodings, the model applies stacked encoder blocks with attention and feed-forward layers. A prepended \texttt{[CLS]} token is used to extract a global representation of the sequence. While the Transformer supports parallel training and models long-range dependencies, its higher complexity may be excessive for our short input windows and limited dataset.

\subsection{Training Procedure and Imbalanced Data Handling} Before training, the segmented dataset was divided into disjoint subsets for model learning and evaluation. We partitioned each participant’s PPG segments into a training set (approximately 60\% of the segments), a validation set (20\%), and a test set (20\%). This per-user stratified split ensures that the model is tested on data from every user that it has seen during training, thereby assessing its ability to recognize individuals under new conditions or at different times. All models were implemented in PyTorch and trained on a customer-grade GPU, so at both training and inference time, the models are lightweight enough to be deployed on mobile or edge devices if needed.

\subsubsection{Class-weighted Loss}
A challenge in training the authentication models was the class imbalance in the data. Although we tried to collect a comparable duration of data from each participant, there are still some users having fewer segments than others. Moreover, in many biometric scenarios, the imposters can be over-represented when combining data from many non-target subjects. To prevent the model from biasing towards classes with more samples, and to avoid creating artificial data through oversampling, we employed a weighted cross-entropy loss during training. In this scheme, each user’s class in the loss function is given a weight inversely proportional to the number of training examples for that user.

Let $C$ be the number of classes and $N_c$ be the number of training samples for class $c$. If $N$ is the total number of training samples, we define the weight for class $c$ as:
\[
w_c = \frac{N}{C \cdot N_c}
\]

These weights are applied to the standard cross-entropy loss, yielding the class-weighted loss:
\[
\mathcal{L}_{\text{weighted}} = -\sum_{i=1}^{N} w_{y_i} \cdot \log p_{y_i}
\]
where $y_i$ is the true class label of sample $i$, and $p_{y_i}$ is the predicted probability for the correct class.

This weighting ensures that an error made on an under-represented user’s data contributes more to the loss than an error on an over-represented user, thereby balancing the influence of each class on the gradient updates. The weighted loss function proved essential for improving the minority-class accuracies and overall system fairness, without noticeably impacting the performance on well-represented classes.

\subsubsection{Hyperparameters and training setup} We trained both the BiLSTM+Attention and Transformer models using the Adam optimizer. The initial learning rate was set to $9.23\times10^{-4}$, and an exponential learning rate decay schedule was applied to promote convergence (the learning rate was decreased by a factor of 0.5 if the validation loss plateaued for a certain number of epochs). We trained for a maximum of 40 epochs for model selection comparison, and 100 epochs for evaluating our methodology on We-Be Dataset. Each training batch consisted of 32 segments. We applied regularization techniques including dropout and weight decay to further mitigate overfitting. Table~\ref{tab:hyperparam_tuning} summarizes the key hyperparameter values and architectural settings for BiLSTM+Attention model.

\begin{table}[ht]
  \centering
  \caption{Hyperparameter search space and optimal configuration}
  \label{tab:hyperparam_tuning}
  \begin{tabular}{lll}
    \hline
    Hyperparameter          & Search Space                   & Optimal Value       \\ \hline
    Hidden dimension        & \{32, 64, 128, 256, 512\}      & 256                 \\
    LSTM layers             & \{1–5\}                        & 3                   \\
    Dropout                 & [0.3–0.5]                      & 0.47                \\
    Learning rate           & $10^{-5}$–$10^{-2}$ (log)      & $9.23\times10^{-4}$ \\
    Weight decay            & $10^{-6}$–$10^{-3}$ (log)      & $8.21\times10^{-6}$ \\
    Epochs (per trial)      & 30                             & —                   \\
    Trials                  & 50                             & —                   \\ \hline
  \end{tabular}
\end{table}

\section{Evaluation}
\label{sec:evaluation}
We evaluate the proposed continuous PPG-based authentication system and present experimental results in this section. We assess model performance in terms of classification accuracy, biometric-specific metrics, learning dynamics, and class-wise behavior, reflecting both system robustness and practical viability. Our evaluation systematically addresses the following research questions:

\begin{itemize}
    \item \textbf{RQ1:} How does PPG sampling frequency affect authentication accuracy and generalization in real-world wearable use cases?
    \item \textbf{RQ2:} To what extent does multi-channel PPG improve authentication resilience against impersonation and signal noise compared to single-channel input?
    \item \textbf{RQ3:} What is the impact of activity-aware training (e.g., walking, typing) on the robustness of PPG-based biometric models in dynamic environments?
    \item \textbf{RQ4:} Can a low-power authentication model operating at 25\,Hz PPG sampling achieve accuracy and robustness comparable to high-frequency counterparts on real-world wearable devices?
    \item \textbf{RQ5:} How feasible is real-time, continuous biometric authentication using low-frequency PPG on embedded devices in terms of power and usability?
\end{itemize}

\subsection{Model Evaluation and Selection}
\label{sec:ModelSelect}
We first compared the BiLSTM+Attention and Transformer models on the PTTPPG dataset using both PPG and ECG inputs. Each model was trained and evaluated under identical preprocessing and data splits.

\begin{figure}[ht!]
    \centering
    \includegraphics[width=\linewidth]{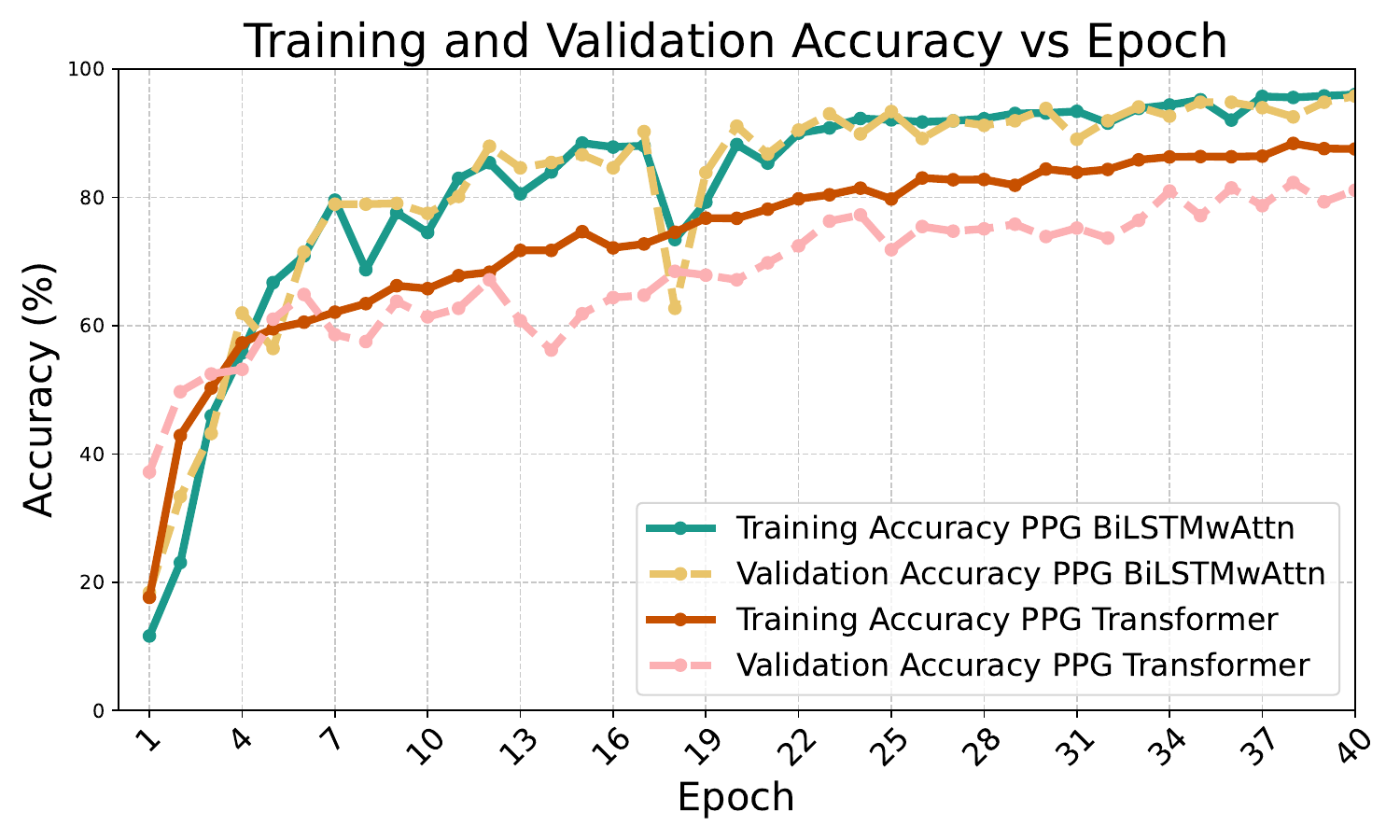}
    \caption{Training and validation accuracy over epochs for the PPG-based models}
    \label{fig:PTTPPG-PPG}
\end{figure}

\begin{figure}[ht!]
    \centering
    \includegraphics[width=\linewidth]{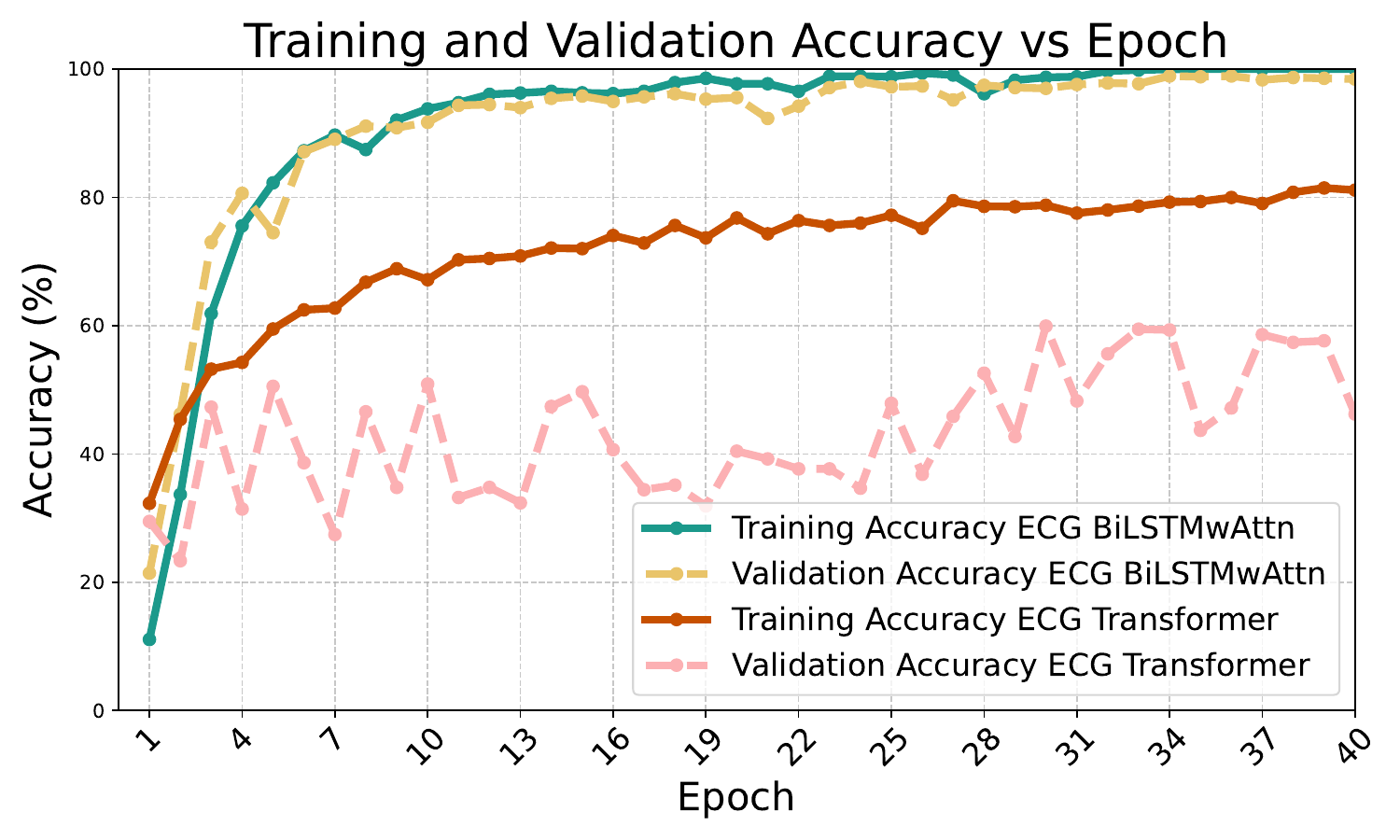}
    \caption{Training and validation accuracy over epochs for the ECG-based models}
    \label{fig:PTTPPG-ECG}
\end{figure}

Figure~\ref{fig:PTTPPG-PPG} shows that both models perform well on PPG signals, with BiLSTM+Attention achieving up to 97\% validation accuracy and outperforming Transformer by several percentage points. In contrast, Figure~\ref{fig:PTTPPG-ECG} demonstrates a dramatic difference in ECG: BiLSTM+Attention maintains strong performance, while the Transformer model fails to converge, showing a 30\% accuracy gap. This suggests that Transformers may be less robust under noisy ECG conditions.

Based on this evaluation, we selected the BiLSTM+Attention architecture as the foundation of our continuous authentication pipeline. Its temporal modeling capacity and generalization make it preferable for deployment on wearable platforms using low-frequency PPG signals.

\subsection{Research Questions (RQs)}
We evaluate our approach to answer the key Research Questions (RQs) described in Section~\ref{RQs} and the beginning of this Section.
\subsubsection{RQ1}
\label{sec:rq1-sampling-rate}

To evaluate the impact of sampling frequency on authentication performance, we conducted an experiment using the PTTPPG dataset. The original PPG signals sampled at 500\,Hz were downsampled to 250\,Hz, 125\,Hz, 25\,Hz, 20\,Hz, 10\,Hz, and 5\,Hz, and we used identical preprocessing and training procedures on each variant. This allowed us to assess performance trends under lower temporal resolutions while holding signal source and model architecture constant. 

\begin{figure}[ht]
    \centering
    \includegraphics[width=\linewidth]{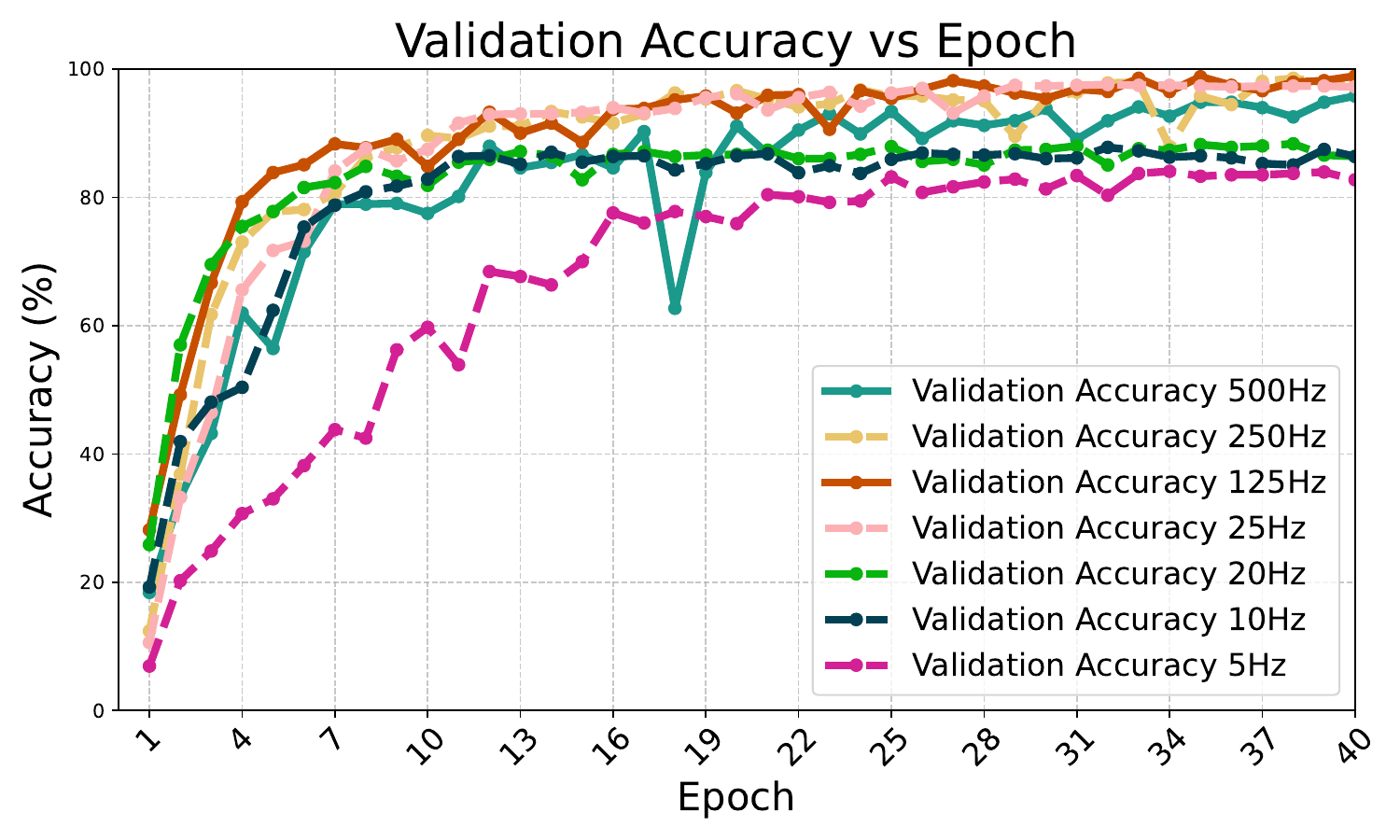}
    \caption{Validation accuracy on PTTPPG dataset under different sampling rates.}
    \label{fig:ppg_downsample_comparison}
\end{figure}

Figure~\ref{fig:ppg_downsample_comparison} presents the validation accuracy curves over 40 epochs. 

Performance remains statistically indistinguishable from the 500\,Hz reference down to 25\,Hz (97–98\,\% peak accuracy) but begins to degrade below that rate. At 20\,Hz, accuracy drops by $\sim$5\,\%, indicating the first tangible loss of discriminatory detail. At 10\,Hz, a similar or slightly larger decline emerges. At 5\,Hz, peak accuracy falls by a further $\sim$8\,\%, confirming that such sparse sampling aliases the pulse waveform beyond reliable recognition; we also observe slower learning during the early epochs.

These results align with the physiological findings from prior work~\cite{choi2017photoplethysmography} that no pulse-rate-variability (PRV) metric differed significantly so long as the sampling frequency was at least 25\,Hz, whereas clear discrepancies appeared at 20\,Hz and below. Together, the evidence suggests that our earlier choice of 25\,Hz—already the default rate on many commercial wearables—hits the sweet spot where power savings plateau yet biometric fidelity is preserved. Rates below 20\,Hz quickly compromise both convergence smoothness and ultimate EER, outweighing any additional energy benefit. A fuller discussion of the physiological and signal-processing implications appears in Section~\ref{sec:discussion-sampling}.

\begin{mdframed}[backgroundcolor=gray!20]
\noindent
\textbf{Conclusion:} Cutting the sampling rate to 25\,Hz preserves authentication accuracy while smoothing convergence and trimming energy use; but dipping below 20\,Hz rapidly degrades both learning speed and discrimination. Hence 25\,Hz is a practical lower bound that balances biometric fidelity with the power constraints of wearable devices.
\end{mdframed}

\subsubsection{RQ2}
\label{sec:rq2-channel}
To evaluate the impact of spectral diversity on biometric modeling, we conducted an ablation study using the We-Be Dataset. Our Bi-LSTM+Attention model was trained in two configurations: (i) a \textbf{single-channel} setting using only one green PPG signal, and (ii) a \textbf{multi-channel} setting incorporating all four available channels (two green, one red, and one infrared).

\begin{figure}[ht]
    \centering
    \includegraphics[width=.9\linewidth]{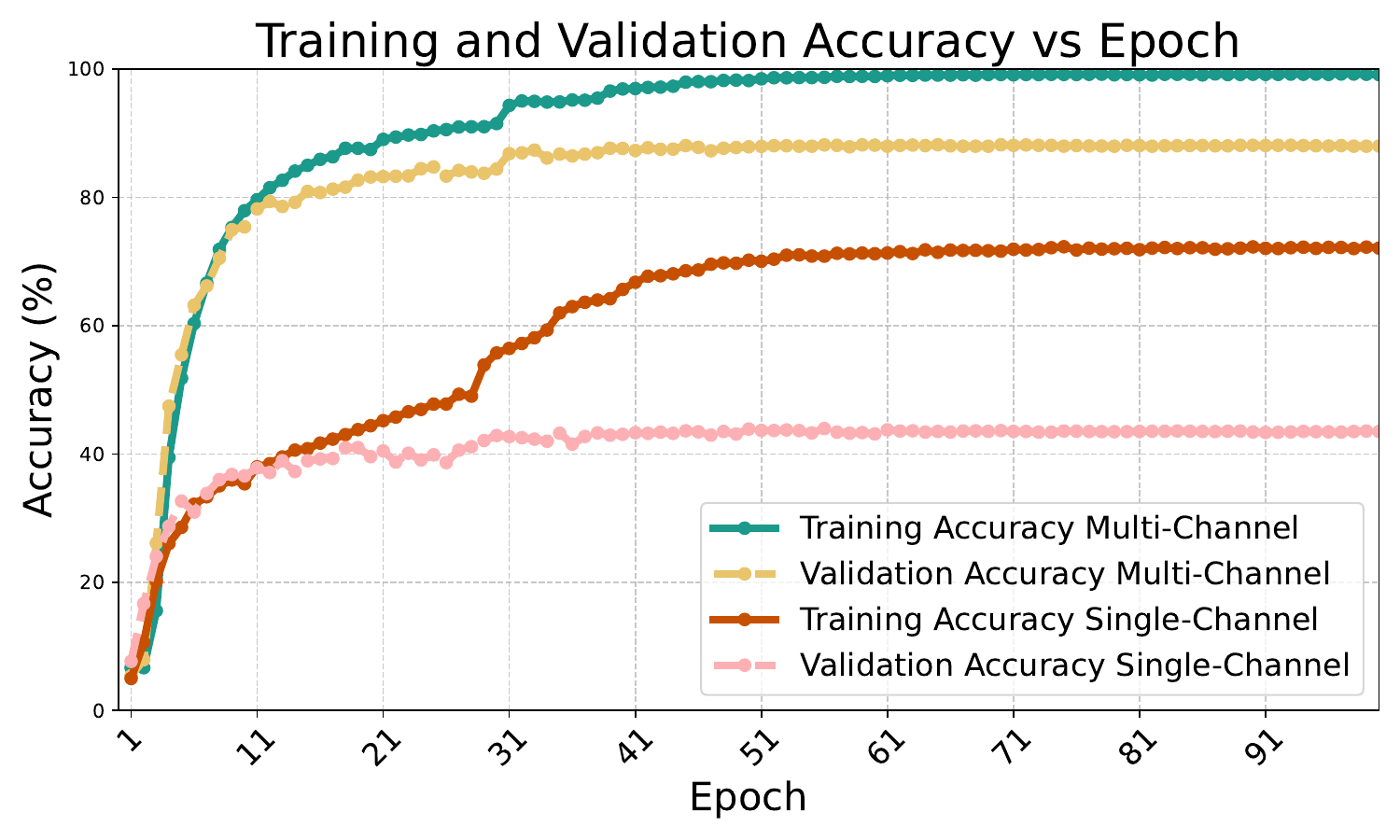}
    \caption{Training and validation accuracy over epochs for single-channel vs. multi-channel PPG input.}
    \label{WEBE-PPG}
\end{figure}

We tracked the training and validation performance of the Bi-LSTM with attention model over 100 epochs using both single-channel and multi-channel PPG inputs on the We-Be Dataset. The resulting accuracy curves are shown in Figure~\ref{WEBE-PPG}.

The multi-channel model exhibited smooth and rapid convergence, reaching a training accuracy of 99.14\% 
and a final validation accuracy of 88.00\% 
These curves suggest stable learning dynamics with modest overfitting, evidenced by the 11\% gap between training and validation accuracy.

In contrast, the single-channel model plateaued at a much lower validation accuracy of approximately 43.5\%, despite steady gains in training accuracy. This large generalization gap implies that single-channel input lacks the diversity and redundancy required to support robust identity recognition, especially under realistic motion and physiological variability. The gap in performance highlights the importance of leveraging multi-channel PPG signals for reliable biometric modeling.

These trends validate the architectural choice of fusing multiple PPG channels and support the decision to avoid minimal-sensing configurations for continuous authentication.

\begin{mdframed}[backgroundcolor=gray!20]
\noindent
\textbf{Conclusion:} Our findings demonstrate that multi-channel PPG significantly enhances the robustness of biometric authentication against noise and inter-user similarity. This supports the adoption of multi-channel sensing in practical deployment, particularly for continuous authentication scenarios where signal quality and user dynamics vary over time.
\end{mdframed}

\subsubsection{RQ3}
\label{sec:rq3-activity-diversity}
To evaluate the impact of activity diversity on model performance, we trained two versions of our Bi-LSTM+Attention model using multi-channel PPG data: one using only resting-state signals, and another using a mix of sitting, walking, and typing activities. Both models were trained on the We-Be Dataset with identical hyperparameters.

\begin{figure}[ht]
    \centering
    \includegraphics[width=.9\linewidth]{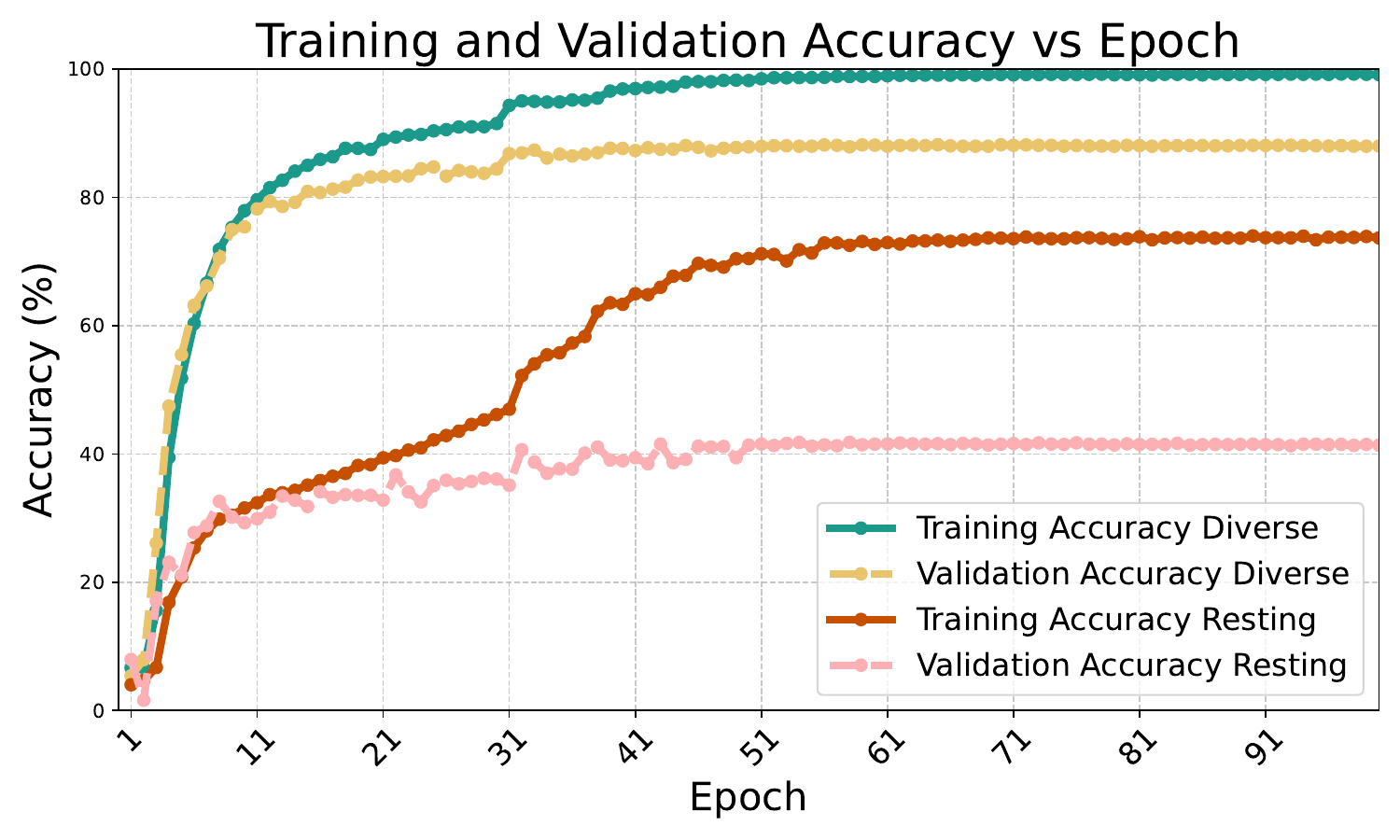}
    \caption{Training and validation accuracy over epochs for models trained with diverse activities vs. resting-only PPG data.}
    \label{fig:resting_vs_diverse}
\end{figure}

As shown in Figure~\ref{fig:resting_vs_diverse}, the model trained on diverse activity data significantly outperformed the resting-only baseline. The diverse model achieved a final validation accuracy of 88.00\%, while the resting-only model plateaued around 41.5\%, despite continued improvements in training accuracy. This large generalization gap in the resting-only model indicates severe overfitting to low-variability heartbeat patterns and a failure to learn robust, identity-specific features. Further discussion on this issue can be found in Section~\ref{sec:activity-diversity-dis}.

\begin{mdframed}[backgroundcolor=gray!20]
\noindent
\textbf{Conclusion:} Training solely on resting PPG data results in poor generalization and should be avoided. Including segments from everyday activities enables the model to learn identity-relevant dynamics that persist across varying physiological states. This strategy is essential for building biometric systems that function reliably in real-world settings.
\end{mdframed}

\subsubsection{RQ4}
\label{sec:rq4-performance}
\begin{itemize}
    \item \textbf{Overall and Class-Wise Performance:} The Bi-LSTM with attention model achieved a test accuracy of 88.11\%. Macro-averaged precision, recall, and F1-score were all 0.88, indicating consistent performance across all classes despite class imbalance. Table~\ref{tab:class_performance} highlights the best- and worst-performing classes, with Class 18 achieving near-perfect accuracy and Class 19 showing relatively lower performance.
    \begin{table}[ht]
      \centering
      \caption{Representative class performance metrics}
      \label{tab:class_performance}
      \begin{tabular}{lcccc}
        \hline
        Class & Precision & Recall & F1-score & Support \\ \hline
        Best (Class 18) & 0.99 & 0.99 & 0.99 & 358 \\
        Worst (Class 19) & 0.76 & 0.68 & 0.72 & 329 \\
        Macro Avg & 0.88 & 0.88 & 0.88 & -- \\ \hline
      \end{tabular}
    \end{table}

    \item \textbf{Authentication Metrics:} Biometric authentication systems are evaluated not just by classification accuracy but also by user-specific error metrics. in our test set, the system achieved:
    \begin{itemize}
      \item False Acceptance Rate (FAR): 0.48\%
      \item False Rejection Rate (FRR): 11.77\%
      \item Equal Error Rate (EER): 2.76\%
    \end{itemize}
    
    These results reflect a practical balance between security and usability: a low FAR ensures that imposters are rarely accpeted, while a moderate FRR could be further mitigated through multi-factor integration or opportunistic gating mechanisms. The ROC curves in Figure~\ref{fig:roc_curve} further support this, with one-vs-all AUC scores consistently above 0.97.

    \begin{figure}[ht]
        \centering
        \includegraphics[width=.9\linewidth]{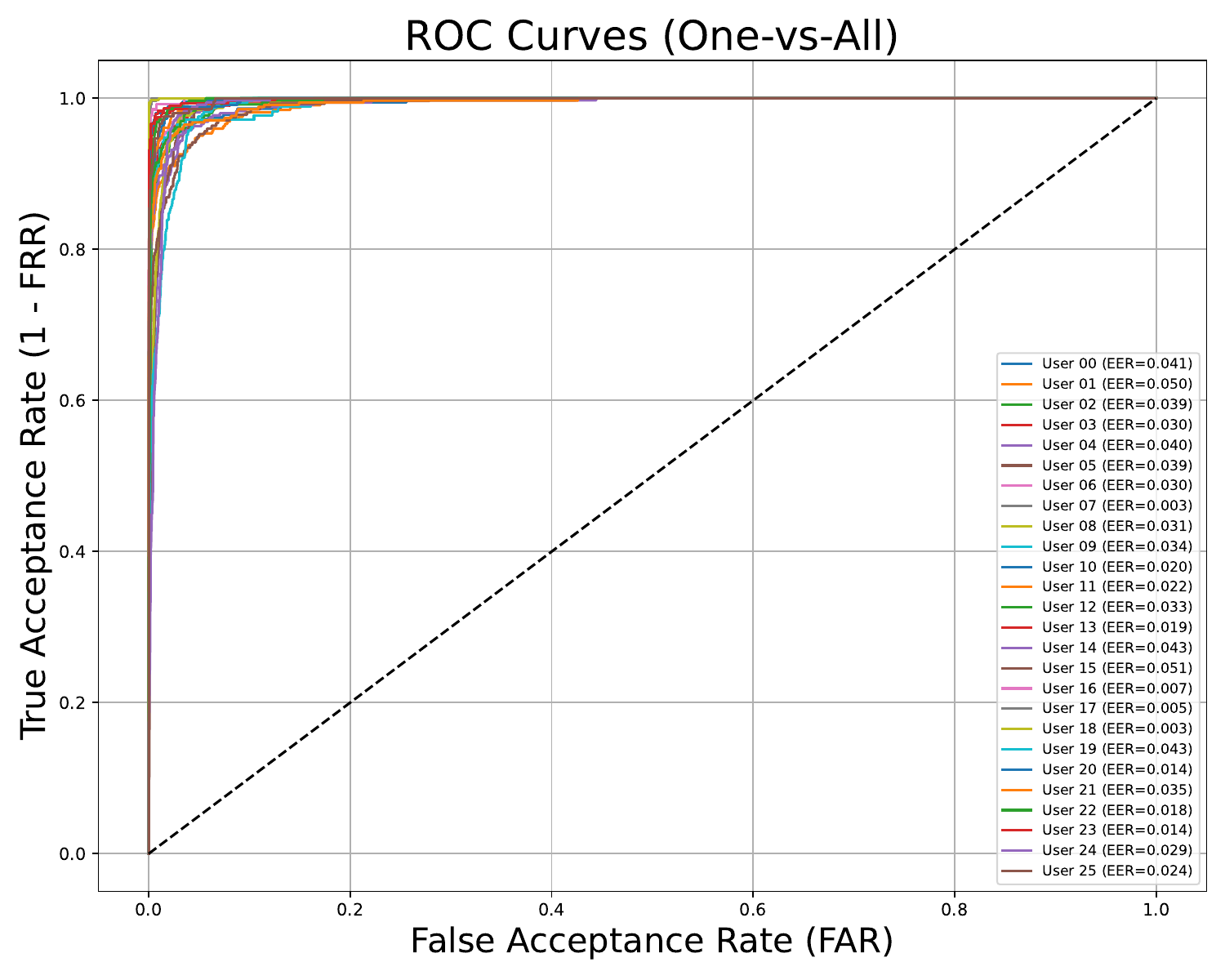}
        \caption{One-vs-All ROC curves for selected users. The system achieves an average AUC above 0.97.}
        \label{fig:roc_curve}
    \end{figure}

    \item \textbf{Confusion Matrix:} The normalized confusion matrix in Figure~\ref{fig:confusion_matrix} is largely diagonal, demonstrating high per-user classification reliability. A few mild off-diagonal errors—such as those between Class 19 and Class 45—are rare and localized, often stemming from users with overlapping physiological patterns or fewer available samples.

    \begin{figure}[ht]
        \centering
        \includegraphics[width=\linewidth]{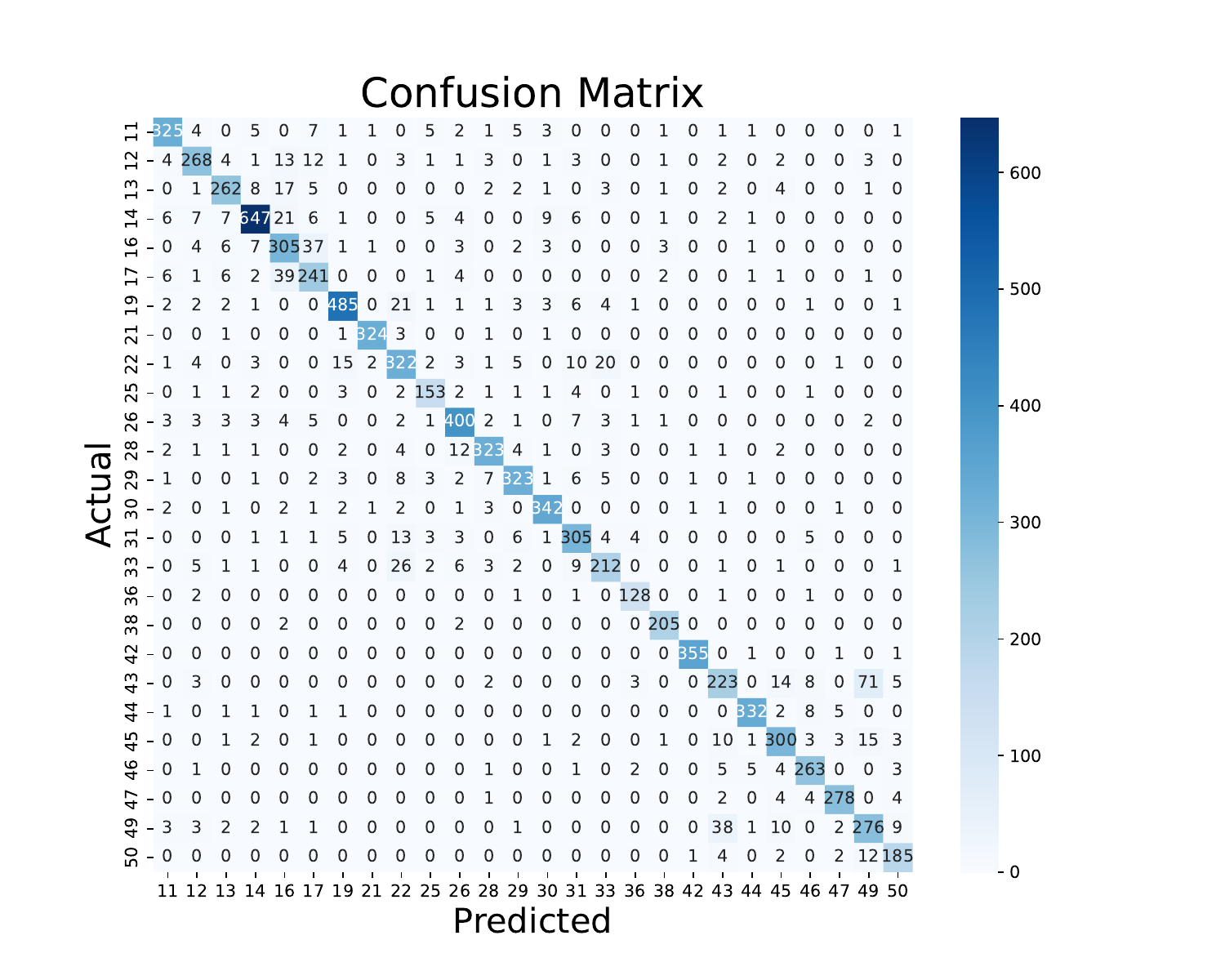}
        \caption{Normalized confusion matrix showing class-wise prediction distribution.}
        \label{fig:confusion_matrix}
    \end{figure}

    \item \textbf{Impact of Class Imbalance:} The number of segments per class ranged from 134 (Class 16) to 723 (Class 3). To counteract this imbalance, we employed class-weighted cross-entropy loss instead of oversampling. This approach preserved physiological signal integrity and led to balanced macro and weighted F1-scores of 0.88.
    
\end{itemize}

\begin{mdframed}[backgroundcolor=gray!20]
\noindent
\textbf{Conclusion:} Our results show that a low-power authentication model operating at 25\,Hz sampling can achieve competitive accuracy and robustness. The high AUC scores and well-balanced per-class metrics underscore the system's real-world applicability despite reduced sampling resolution. By maintaining low FAR and mitigating class imbalance through a class-weighted loss, our approach demonstrates the feasibility of deploying lightweight biometric models on resource-constrained wearable devices without sacrificing authentication integrity.
\end{mdframed}

\subsubsection{RQ5}
\label{sec:rq5-power-consumption}

\begin{figure}[ht]
    \centering
    \includegraphics[width=.9\linewidth]{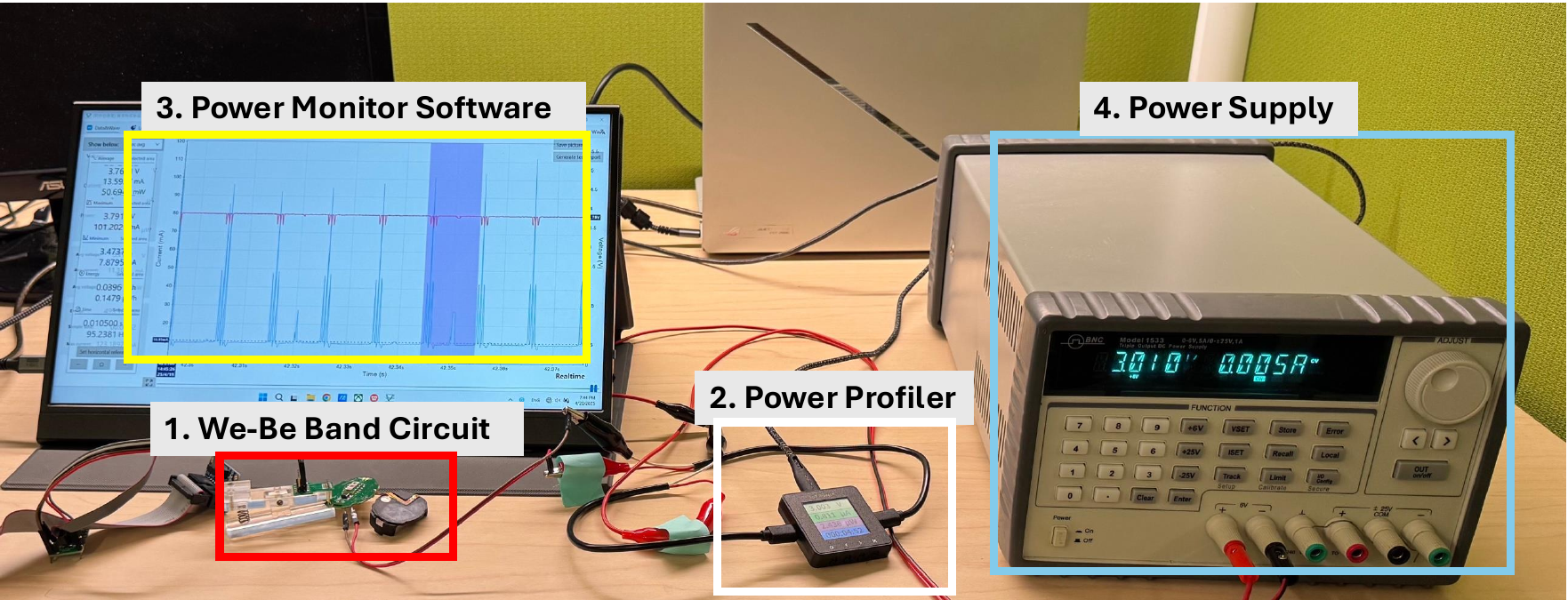}
    \caption{Experimental setup to measure the power consumption of the PPG sensing module.}
    \label{FigPowerSetup}
\end{figure}

To quantify the energy implications of operating our authentication system at different sampling rates, we directly measured the power consumption of our PPG sensor while varying the sampling rate from 8\,Hz to 512\,Hz. The experimental setup is shown in Figure~\ref{FigPowerSetup}. We dismantle a We-Be band and expose the circuit board, which is connected to a power profiler to obtain power consumption data.

\begin{figure}[ht]
    \centering
    \includegraphics[width=\linewidth]{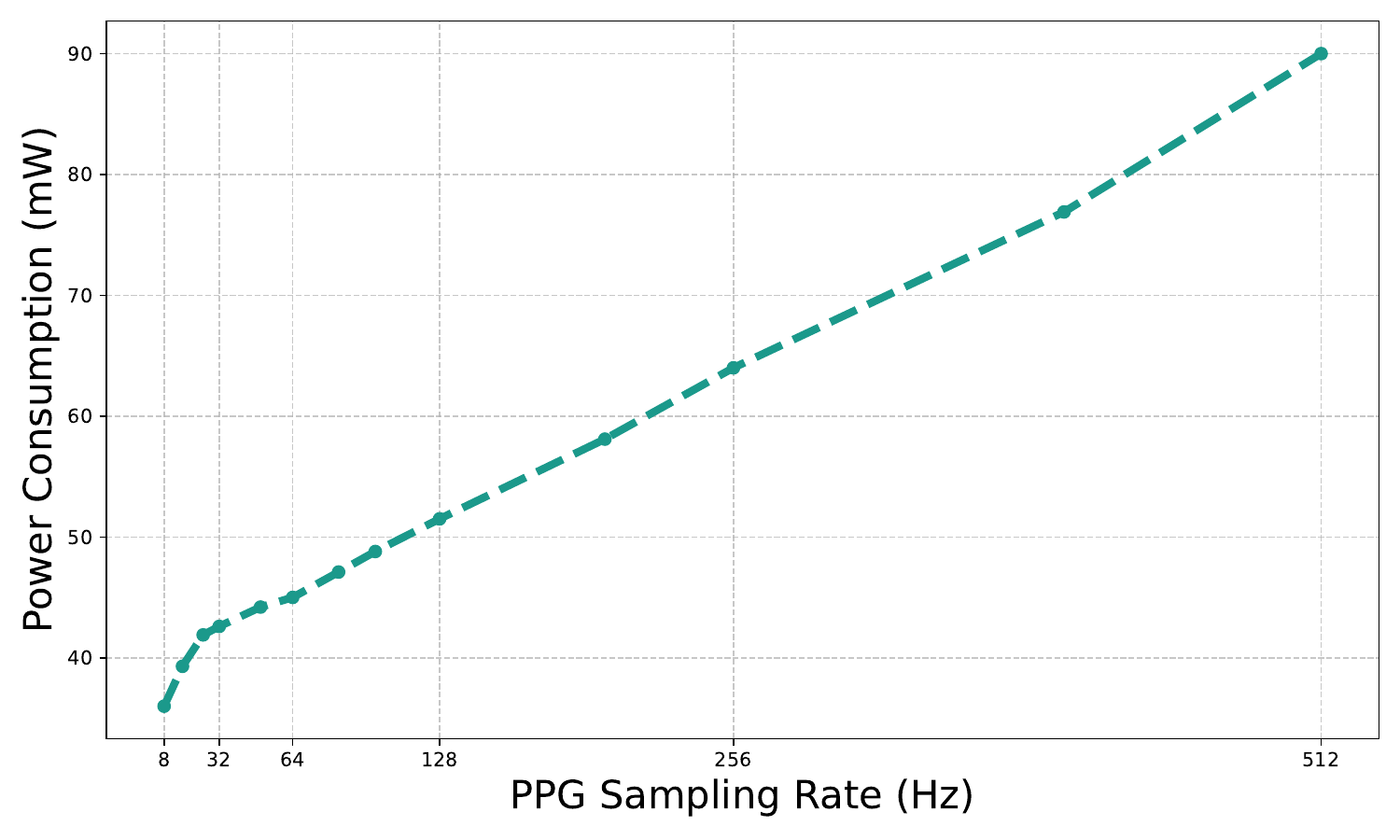}
    \caption{Sensor power consumption as a function of PPG sampling rate.}
    \label{FigSensorPower}
\end{figure}

As illustrated in Figure~\ref{FigSensorPower}, power consumption increases nonlinearly with sampling frequency. Below 100\,Hz, energy cost scales moderately with frequency; however, above this point, the marginal power draw rises steeply. This behavior is likely due to increased LED duty cycles and sensor controller overhead at higher sampling rates.

While prior studies have demonstrated excellent performance using datasets collected at sampling rates of 75\,Hz or higher, our results show that such high rates are not strictly necessary for effective continuous user authentication. We find that a significantly lower rate of 25\,Hz can yield comparable authentication performance.

At this low sampling rate, the PPG module consumes just 41.9\,mW, compared to 51.5\,mW at 128\,Hz and 90.0\,mW at 512\,Hz. These savings are especially significant in the context of battery-powered wearable devices, where sensing accounts for a large fraction of energy usage.

\begin{mdframed}[backgroundcolor=gray!20]
\noindent
\textbf{Conclusion:} Operating our system at 25\,Hz yields substantial energy savings—up to 53\% lower than 512\,Hz and 19\% lower than 128\,Hz—without compromising authentication performance. This validates the practicality of deploying our framework on ultra-low-power wearable platforms, extending operational lifespan without sacrificing biometric accuracy.
\end{mdframed}

\subsection{Realistic Long‑Term Usage and Battery Life on the We‑Be Band}
\label{sec:long-term}

To test real‑world viability we conducted an independent ten‑day realistic long-term study with one participant wearing a fully charged We‑Be Band that streamed four‑channel PPG at 25\,Hz. The study followed three phases:
\begin{enumerate}
  \item \textbf{Training data collection:} For ten consecutive days the participant recorded 20 min of PPG while resting, typing, and walking, mirroring the activity mix used in our earlier evaluation.
  \item \textbf{Continuous run for battery lifetime and testing data collection:} After the tenth day the watch was fully charged and worn continuously until it dies, streaming PPG data throughout daily life with more diverse and zero-shot activities like driving, stairs, strength workouts, dining, and watching TVs, and widely varying ambient light across different environment and day time, making it more challenging.
  \item \textbf{Incremental training and evaluation:} Ten models were trained, each adding one extra day of data. A single 18 h “awake‑only” slice, excluding the sleep period, from the testing data served as the test set.
\end{enumerate}

Continuous sensing exhausted the battery in 1,550\,minutes (25.8\,hours), confirming that the We‑Be Band can support day‑long 25\,Hz PPG collection on a single charge.

\begin{figure}[ht]
    \centering
    \includegraphics[width=\linewidth]{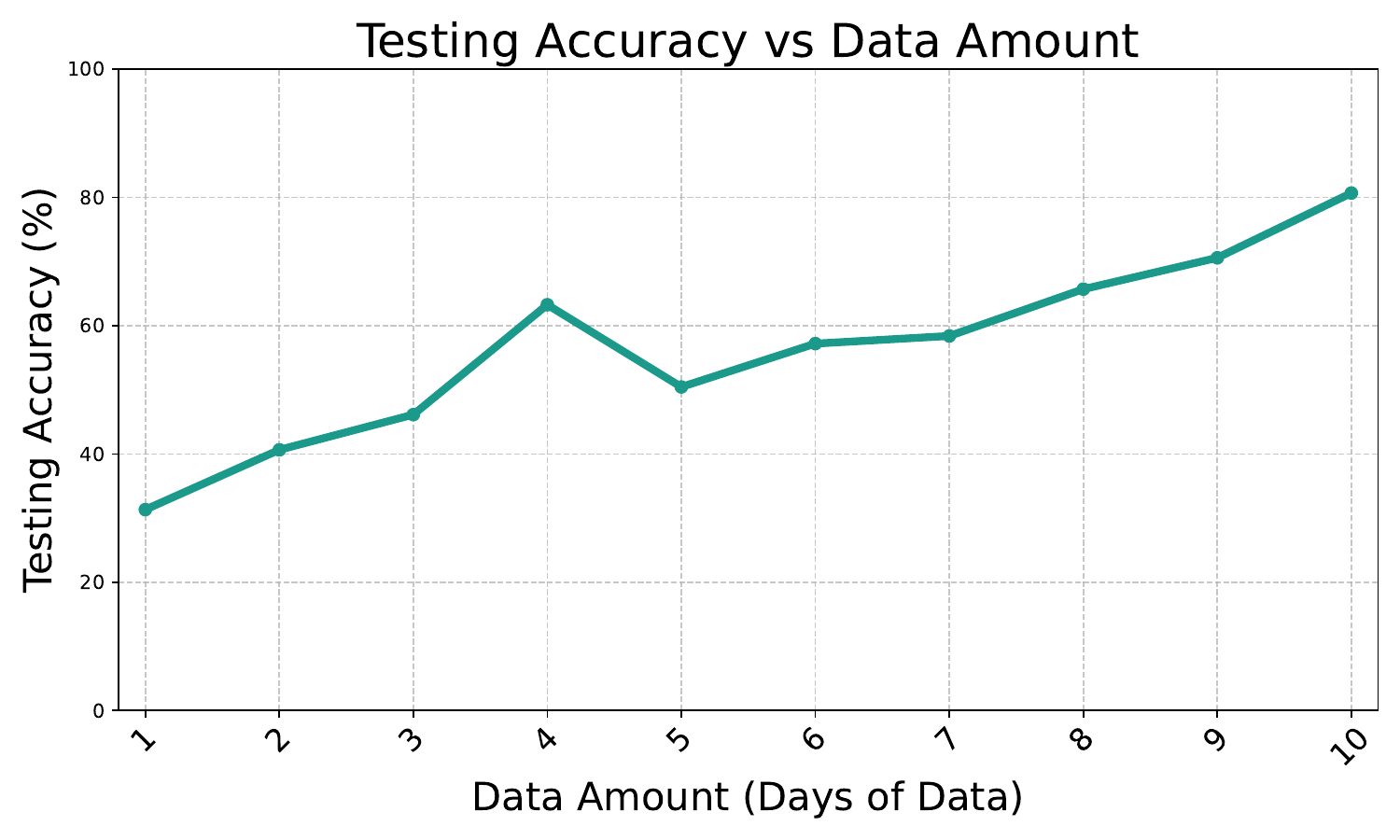}
    \caption{Test accuracy on an 18 h “awake” segment versus the number
    of days' data used for training.}
    \label{Fig:Long-term}
\end{figure}

Figure~\ref{Fig:Long-term} plots test accuracy over different data amount. Accuracy climbs steadily, reaching 80.7\% with all ten days—about 8\% below the previous benchmark. The drop is expected because the long‑term test (i) spans a much longer duration, (ii) includes unfamiliar and higher‑intensity activities such as workouts in gym, driving, and going up and down stairs, and (iii) faces changing indoor/outdoor lighting. Even so, just ten days of short daily sessions ($<$3.5 h total PPG) already deliver reliable, day‑long authentication while the watch lasts a full battery cycle. With continued incremental fine‑tuning and the gradual accumulation of varied high‑quality data, we expect the long‑term accuracy to converge toward—if not match—the 88.11\% achieved in our previous evaluation, confirming the practical value and headroom of the proposed approach.




\section{Prototype Evaluation}
\label{sec:case-study}

We implemented our continuous PPG-based authentication system based on the We-Be Band wearable device and evaluated its real-time performance in a real-world usage scenario involving 20 users. The goal was to determine whether our approach can effectively differentiate between a legitimate user and unknown users (or imposters) under realistic operating conditions.

The usage of such a device involves several stages:
\begin{enumerate}
    \item \textbf{Data collection:} where PPG data belonging to the target user is collected;
    \item \textbf{Model training:} where a model is trained to recognize a user based on collected PPG data;
    \item \textbf{Real-time continuous authentication:} where the trained model starts to work and identify if the device is being operated by the corrected user.
\end{enumerate}

\subsection{Experiment Setup}
\subsubsection{Hardware}
\begin{figure}[ht!]
    \centering
    \includegraphics[width=.6\linewidth]{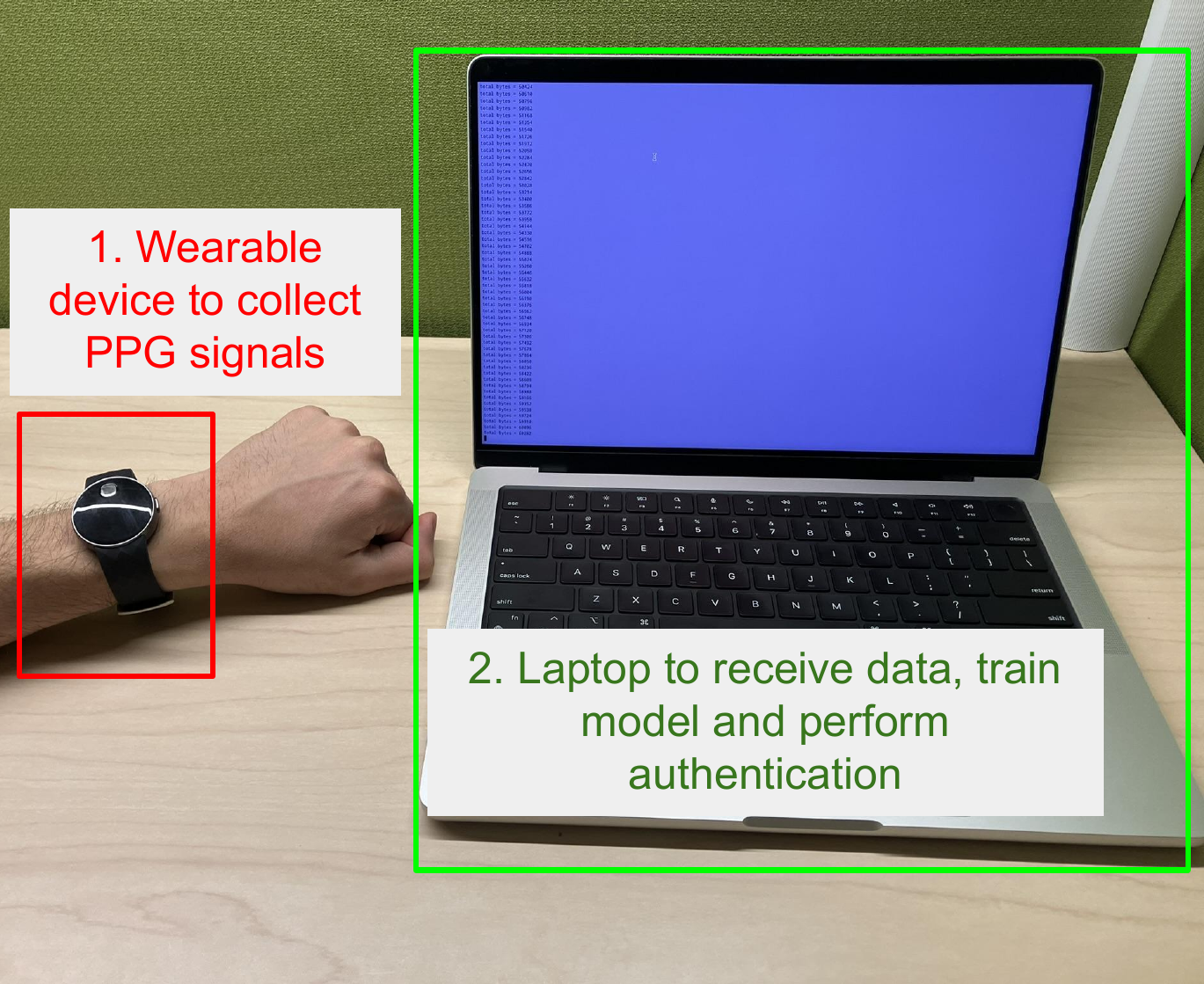}
    \caption{Case study hardware setup.}
    \label{FigDemoExp}
\end{figure}

The We-Be Band wearable device was used for PPG signal acquisition. A MacBook Pro laptop equipped with an Apple M-series processor serves as the gateway device for receiving PPG data, training models, and performing inference. We choose to perform computation tasks on a laptop instead of using cloud or more powerful servers because we want to show that all the tasks can be done on resource-constraint devices, and the whole workflow can be easily ported to mobile devices with similar computing powers. The We-Be Band and laptop communicated via Bluetooth Low Energy (BLE). A picture of our experiment setup is shown in Figure~\ref{FigDemoExp}.

\subsubsection{Experiment Process}

\begin{figure}[ht!]
    \centering
    \includegraphics[width=.9\linewidth]{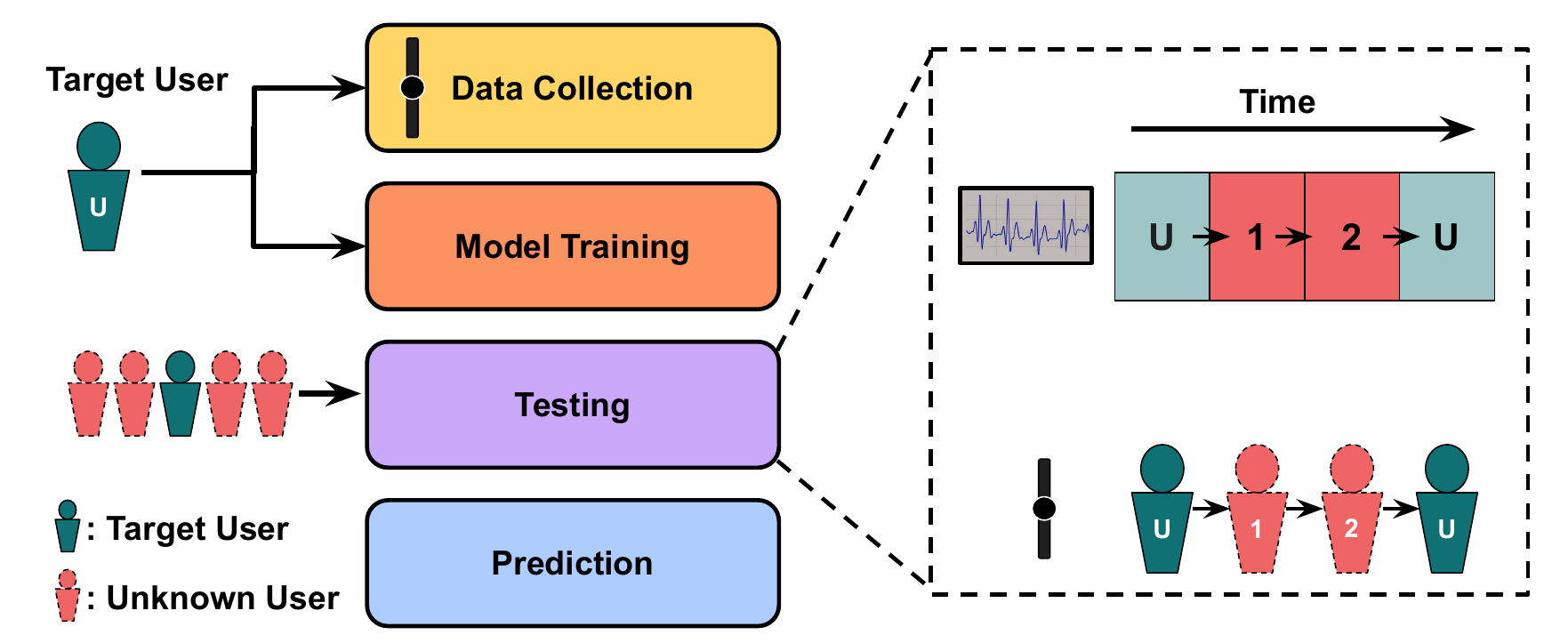}
    \caption{Experiment steps.}
    \label{FigCaseStudyDiagram}
\end{figure}

The flow of our experiments is shown in Figure~\ref{FigCaseStudyDiagram}. This experiment involves 20 different individuals. Each user follows a fixed timeline: 
\begin{enumerate}
    \item \textbf{Data collection:} The user wears a watch for around 55 minutes to collect 30,000 data points.
    \item \textbf{Model training:} A dedicated model is trained based on the newly collected data, mixed with the We-Be Dataset. After this step, the user will have a dedicated model labeled with their name.
    \item \textbf{Testing:} We follow a fixed routine to switch the watch between different users. First, after booting the watch, the user wears the watch for 4.5 minutes. Then the watch is handed to 2 ``imposters'' whose PPG signals are not included in the model training set. Each imposter user wears the watch for 2.5 minutes. Finally, the user wears the watch again for another 2.5 minutes.
    \item \textbf{Data Collection:} During Step 3), real-time authentication decisions were generated and logged for further analysis.
\end{enumerate}

The time it takes for each experiment is shown in Table~\ref{TabCaseStudyTime}. The combined data collection and model training time can also represent the time it costs to set up a watch for a new user in a real-world scenario.

\begin{table}[ht]
  \centering
  \caption{Time Span.}
  \label{TabCaseStudyTime}
  \begin{tabular}{ccc}
    \hline
    Data Collection   & Model Training  &  Testing    \\ \hline
    55 min            & 49 min          & 4.5 + 2.5 + 2.5 + 2.5 = 12 min       \\  
    \hline
  \end{tabular}
\end{table}

\subsection{Results}
\subsubsection{Raw vs. Smoothed Predictions}
\begin{figure*}[ht!]
    \centering
    \begin{subfigure}[t]{.45\linewidth}
         \centering
         \includegraphics[width=\linewidth]{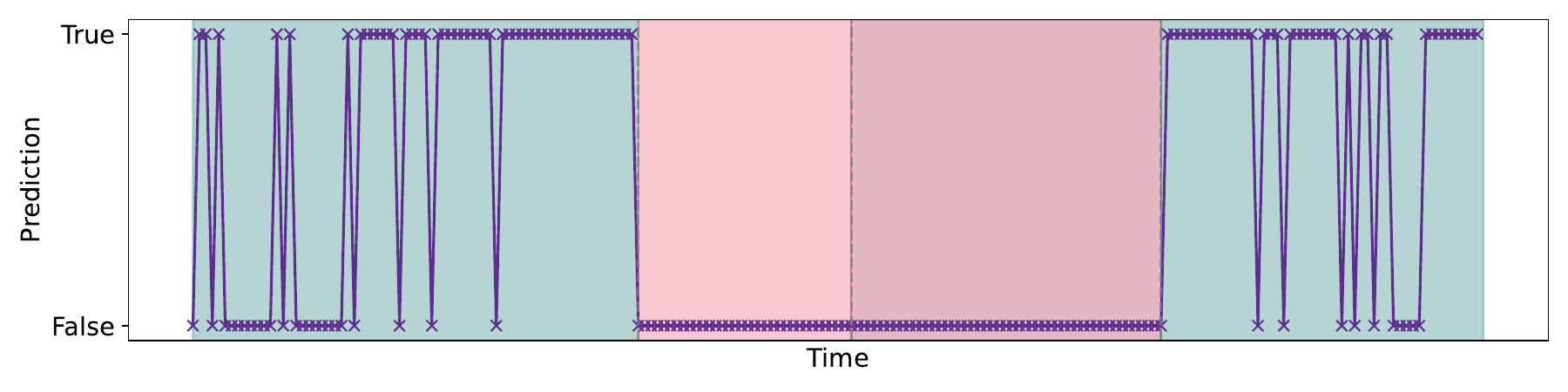}
         \caption{User 1.}
    \end{subfigure}
    \centering
    \begin{subfigure}[t]{.45\linewidth}
         \centering
         \includegraphics[width=\linewidth]{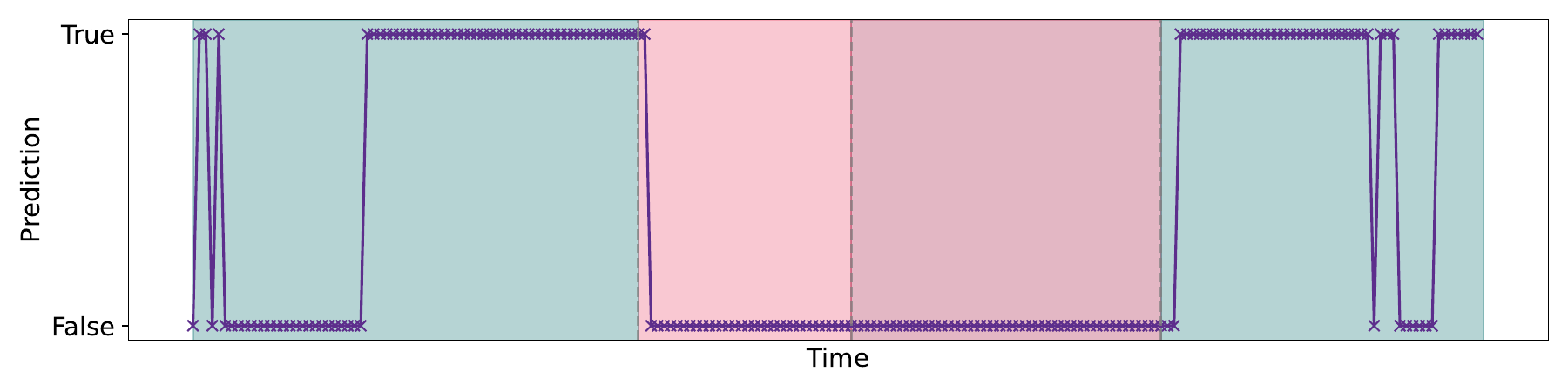}
         \caption{User 1, with a sliding-window filter applied.}
    \end{subfigure}
    \centering
    \begin{subfigure}[t]{.45\linewidth}
         \centering
         \includegraphics[width=\linewidth]{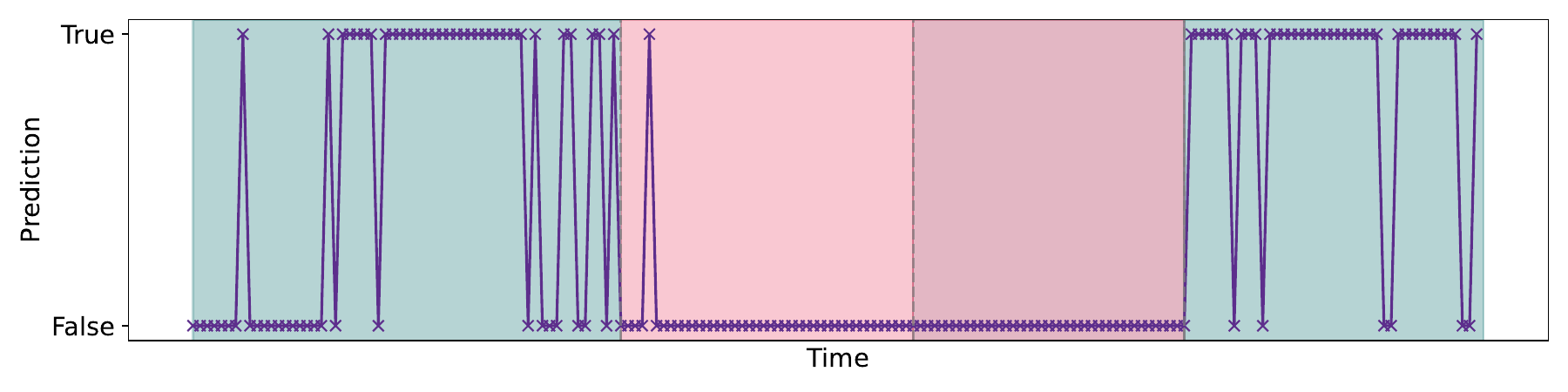}
         \caption{User 2.}
    \end{subfigure}
    \centering
    \begin{subfigure}[t]{.45\linewidth}
         \centering
         \includegraphics[width=\linewidth]{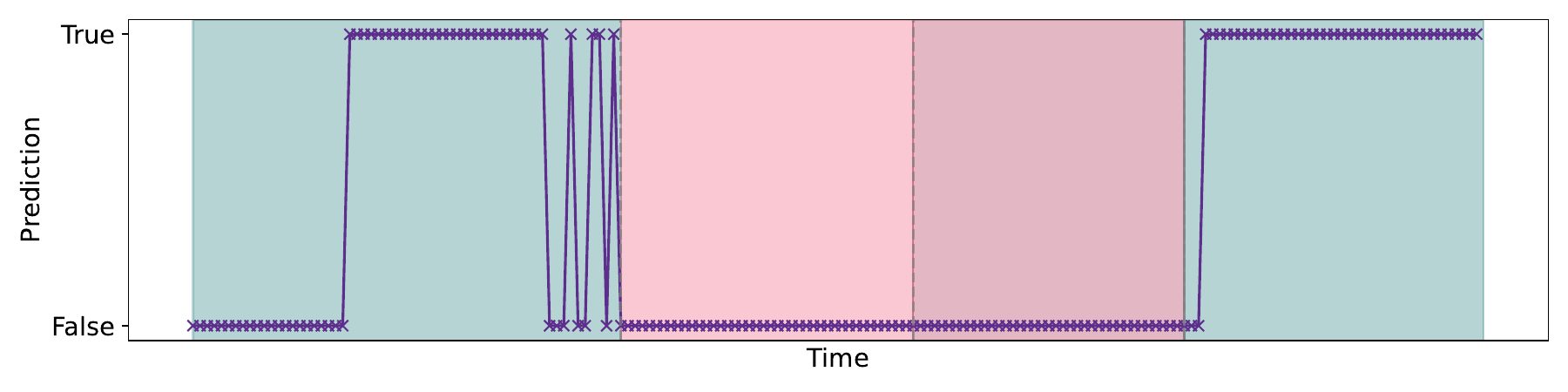}
         \caption{User 2, with a sliding-window filter applied.}
    \end{subfigure}
    \centering
    \begin{subfigure}[t]{.45\linewidth}
         \centering
         \includegraphics[width=\linewidth]{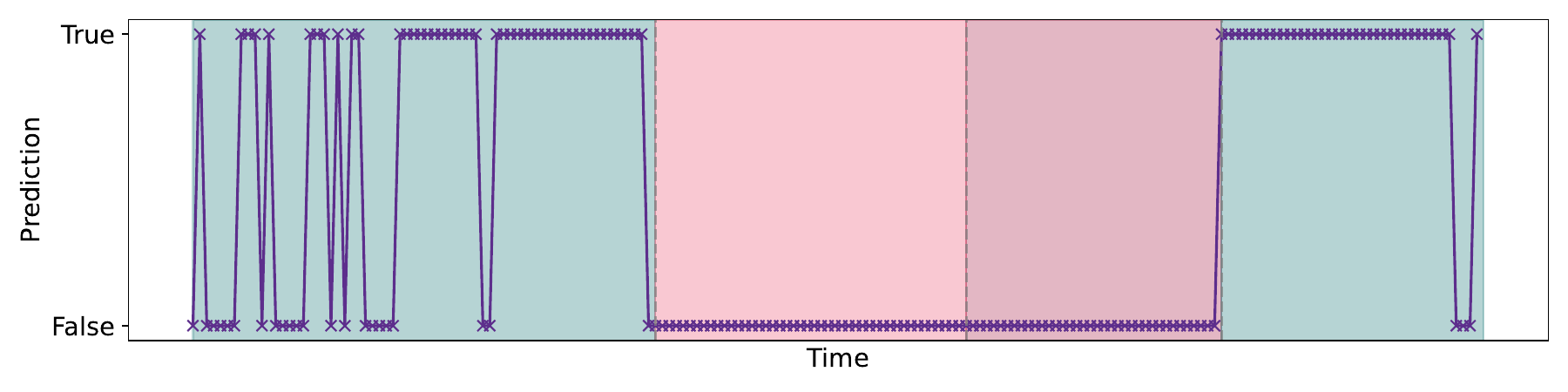}
         \caption{User 3.}
    \end{subfigure}
    \centering
    \begin{subfigure}[t]{.45\linewidth}
         \centering
         \includegraphics[width=\linewidth]{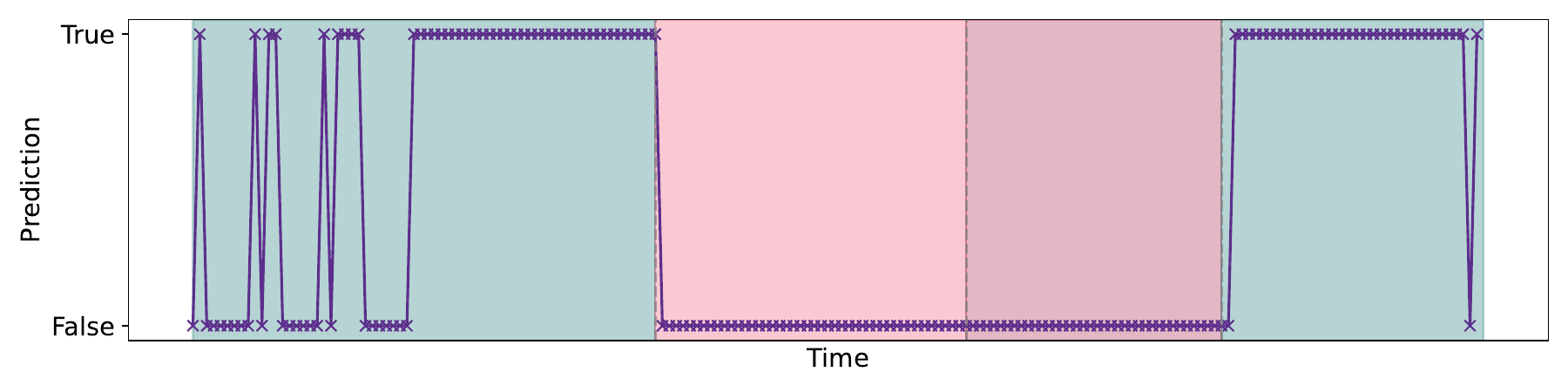}
         \caption{User 3, with a sliding-window filter applied.}
    \end{subfigure}
    
    \caption{Real-time authentication results on 3 participants. The green background color indicates that the watch is worn by a legitimate user, while the red background colors indicate that the watch is worn by imposters.}
    \label{FigDemoResult}
\end{figure*}

Our results on every individual show that the authentication system works for every test subject. We select to show the generated prediction traces of 3 users in Figure~\ref{FigDemoResult}. We can see that in raw data, despite being correct most of the time, the model outputs brief false rejections occasionally. To address this, we apply a sliding window filter to the trace of prediction results, where the current values are determined by previously observed values within a fixed-size window, and are replaced by the majority class in that window when a clear majority exists. Applying a sliding window majority filter significantly reduces prediction jitter, resulting in stable authentication streams. We can see that our authentication system is reliable from the traces shown. Excluding the device warm-up phase, where the We-Be Band spends around 2 minutes to calibrate its sensors and internal DSP algorithm parameters, our method completely rejects all unauthorized access to the device, and successfully verifies the correct user most of the time, although with occasional jitters.

\subsubsection{Handling Unknown Users}
During inference the model assigns an unknown label whenever the maximum soft‑max probability falls below the empirically chosen 0.8 threshold; this open‑set trigger is then smoothed with the same 5‑sample majority filter used for genuine decisions, yielding the practical result shown above. These observations demonstrate that the prototype can reliably reject users whose PPG was unseen during training.

\subsubsection{Deployment Insights}
These results confirm the feasibility of deploying our system in real-world settings. The model (1) Quickly detects imposter transitions ($<$5s delay), (2) Maintains continuous recognition of genuine users, and (3) Perfectly rejects all imposters.

To the best of our knowledge, this is the first work to fully implement and evaluate a continuous biometric authentication system on a smartwatch.

These findings establish the viability of continuous authentication using low-frequency PPG and pave the way for future deployment in smartwatch, consumer- and medical-grade wearable systems.

\section{Discussion}

\subsection{Why Did Some Down Sampled Data Perform Similar as Original Data in PTTPPG Dataset Experiment?}
\label{sec:discussion-sampling}

In Section~\ref{sec:rq1-sampling-rate}, our Bi-LSTM+Attention model trained on some downsampled PPG data performed similar as models trained on the original 500\,Hz signals from the PTTPPG dataset. While this may seem counterintuitive—given the assumption that higher-frequency signals contain richer physiological detail—there are important caveats and potential explanations:
\begin{itemize}
  \item \textbf{Regularization effect:} Downsampling can act as a form of regularization, reducing high-frequency noise or variability that is irrelevant to biometric identity.
  \item \textbf{Feature sparsity:} PPG signals are quasi-periodic. Oversampling can lead to redundant temporal features, while moderate downsampling preserves essential waveform morphology without overwhelming the model.
  \item \textbf{Model fit:} High-dimensional input from high-frequency signals can exceed the capacity of compact models, particularly in settings with limited training data.
\end{itemize}


\subsection{Applicability to Other Devices}
\label{sec:cross-device} 
Our evaluation assumed a single, fixed smartwatch per user. In practice, users may replace or upgrade their wearable, and even nominally identical hardware can differ in LED spectrum, photodiode gain, or skin coupling. We therefore expect a degree of device specificity: a model enrolled on one watch may not transfer perfectly to another. Analogous to today’s onboarding flow for commercial biometrics, where a short calibration is required whenever the user pairs a new device. Future work should explore rapid calibration strategies—e.g., few-minute fine-tuning, or domain-adversarial layers—to minimize user effort while coping with sensor drift.

\subsection{On-Device Training and Personalization}
\label{sec:ondevice-train} 
Our prototype streams PPG to a laptop for convenience, but three factors suggest that full edge deployment is feasible: 
\begin{enumerate}
    \item \textbf{Model size:} Modern microcontrollers and smartwatch chipsets now ship with accelerators and on-chip memory measured in the megabytes, more than enough for the lightweight recurrent network used here. 
    \item \textbf{Incremental learning:} Poetntial personalization can perform
      nightly on‑device in four steps:

      \begin{enumerate}
        \item \textit{Clean‑window selection:}  
              A lightweight Signal Quality Assessment (SQA) filter writes only high‑quality segments (4 s, 50\% overlap) from the past 24 h to a rolling buffer capped at 5 min (\(<\!3\) kB, 8‑bit).
        \item \textit{Few‑shot fine‑tuning:}
              At the next charging event—or after midnight if the watch is not worn, update only the projection layer and the second BiLSTM gate ($\sim$8\% of the weights).
        \item \textit{Regularization: }
              An \(L_2\) penalty for weight drift plus elastic weight consolidation for shared layers prevents catastrophic forgetting while allowing gradual adaptation to changes in the skin or sensor.
        \item \textit{Rollback safeguard:} The watch keeps the previous weight snapshot; if the new model’s confidence falls below 80\%, for example, during the following day, it automatically reverts and flags the update for review in the companion app.
      \end{enumerate}

    \item \textbf{Energy budget:} Opportunistic, event-driven authentication can keep duty cycle low so that occasional fine-tuning adds negligible battery cost relative to the watch’s daily charge pattern. 
\end{enumerate}
A fully self-contained implementation would also eliminate the BLE data path, further reducing privacy risk.

\subsection{Security and Privacy Considerations}
\label{sec:security} 
\paragraph*{Spoofing / Presentation Attacks} Replicating a live, multichannel PPG signal would likely require a custom prosthetic or micro-fluidic “arm” capable of reproducing the target’s haemodynamic response in real time—far beyond current commodity threat models. Nevertheless, we outline three hardening layers that collectively make such attacks even less practical:
\begin{enumerate}
    \item \textbf{Random LED pattern:} Each 4 s window uses a pseudo‑random sequence of LED intensity or colors (channels). The pattern is revealed only after the window closes, so a replay device cannot anticipate which optical channel to fake. Any mismatched channel lowers the similarity score and triggers an “unknown” label.
    \item  \textbf{Cross-checking inertial data for liveness:} Every real pulse causes tiny, natural wrist motions that the built‑in accelerometer can feel. We can compute the magnitude‑squared coherence between the radial‑axis accelerometer and the PPG magnitude spectrum at a certain frequency. Small coherence below a certain threshold suggests the optical signal is decoupled from physical motion, a red flag for prosthetic injection.
    \item \textbf{One‑class pulse‑shape anomaly detector:} A lightweight auto‑encoder, trained only on the existing user’s clean pulses, reconstructs each beat at inference time. Reconstruction error above certain threshold marks the segment as spoof‑suspect. Because the detector learns fine‑grained morphology, even perfectly timed yet physiologically unnatural waveforms are rejected.
\end{enumerate}

\paragraph*{Adversarial Machine-Learning Threats} Synthetic perturbations or generative signals might be crafted to fool the classifier. Although a full robustness analysis is outside our scope, common defenses include adversarial training on transformed windows, frequency-domain masking, and score-level temporal smoothing~\cite{shao2022defending,bo2022pretraining,zhao2020trueheart}.
\paragraph*{Privacy Leakage} In our prototype, we stream raw PPG over BLE only to a trusted laptop in the lab; no biometric profile is stored on the watch, and neither the laptop or the watch stores raw data. Combined with the limited access to query the authentication ML model, we believe the privacy risk is minimal. A production system could perform any aggregation using federated or split learning with differential-privacy noise, thereby limiting the risk of privacy leakage.
\paragraph*{Availability Versus Sudden Changes} Majority-vote smoothing across several windows prevents a single outlier—caused by stress, motion, or bad contact—from locking the device. More advanced adaptive thresholds could further balance usability and security, particularly during rapid physiological shifts (e.g., exercise to rest).

\subsection{Effect of Activity Diversity on Model Generalization}
\label{sec:activity-diversity-dis}

Traditionally, resting-state physiological signals are considered optimal for biometric authentication due to their stability and low noise. However, our results reveal a counterintuitive finding: models trained solely on resting PPG segments performed significantly worse when evaluated on signals recorded during walking or other physical activities. Specifically, while the Bi-LSTM+Attention model achieved decent accuracy on held-out rest segments, its performance dropped significantly on other segments from the same users—highlighting a failure to generalize beyond static conditions.

To address this, we introduced activity diversity into the training data, incorporating segments from sitting, walking, and typing, where these mild activities introduce minimal motion artifacts. This broader training regime encouraged the model to extract identity-specific features that remain stable across physiological states, such as inter-channel timing shifts, amplitude morphology, and pulse modulation patterns. 

\begin{itemize}
    \item \textbf{Physiological Insight:} Resting PPG signals often exhibit low inter-subject variability due to stabilized autonomic activity, resulting in homogeneous waveform structures that limit biometric separability. In contrast, physical activity induces sympathetic activation, heart rate elevation, and dynamic vascular responses—all of which vary from person to person. These activity-driven variations act as “physiological fingerprints,” revealing robust identity-linked patterns masked during rest.
    \item \textbf{Real-World Validation:} The benefits of activity-diverse training extended beyond offline metrics. In live deployment on the We-Be Band, users experienced smooth authentication across sitting, walking, and typing activities, demonstrating the system's robustness in everyday use cases. The model retained high accuracy without explicit motion classifiers or condition-specific tuning.
\end{itemize}

Importantly, this variation is not merely noise—it reflects meaningful, individualized cardiovascular responses. While motion artifacts do exist, deep learning models can disentangle such artifacts from systematic, identity-relevant features embedded in the physiological response.

Contrary to conventional wisdom, resting PPG alone is insufficient for reliable authentication. Incorporating activity diversity into the training data is essential to avoid overfitting and ensure generalization. Our findings suggest that mild physiological perturbation—such as walking—can reveal identity-specific dynamics and improve model robustness, even in the presence of moderate signal variability.

\subsection{Device Variability and Calibration}
\label{sec:device-variability}

During evaluation, we observed a strong dependency between the authentication model and the specific device used during data collection. That is, a model trained on PPG data from one smartwatch does not generalize well to data collected from a different unit, even of the same model. This device variability likely stems from subtle hardware-level differences such as sensor alignment, optical path, LED intensity, and skin contact geometry.

While this device-specificity might appear to limit generalization, we argue that it aligns well with our system’s intended use case. Our goal is to link a specific user to a specific device in a continuous authentication scenario. In practical deployments, the authentication model is expected to run on the same wearable device that the user initially sets up. This mirrors the onboarding process of commercial biometric systems, where calibration or enrollment is performed on the user’s own device.

We treat the device-model pairing as part of the personalization process. If a user changes devices, a one-time re-enrollment or calibration stage can be triggered. This auto-calibration step can involve collecting a short segment of resting-state PPG to fine-tune or reinitialize the model.

While the current approach assumes static device usage, future work could investigate transfer learning or domain adaptation techniques to improve cross-device generalization without full retraining.

\section{Related Work}
\label{sec:related_work}

\subsection{PPG-Based Biometric Authentication}
Photoplethysmographic (PPG) signals have emerged as a promising biometric modality due to their accessibility and compatibility with wearable devices~\cite{8411233,6004938,6656145}. Early work by~\cite{8411233} demonstrated robustness across physiological states (e.g., exercise, emotional stress) using Continuous Wavelet Transform (CWT) and Direct Linear Discriminant Analysis (DLDA), achieving equal error rates (EER) of 0.5\%--6\% with short training times. Similarly,~\cite{6004938} validated PPG’s feasibility for identification under controlled conditions, while~\cite{6656145} explored continuous authentication via long-term monitoring. However, these studies relied on high sampling rates ($\geq$50\,Hz) and computationally intensive preprocessing (e.g., CWT), neglecting energy constraints critical for wearable deployment in clinical settings like remote patient monitoring.

\subsection{Deep Learning for PPG Authentication}
Recent advances leverage deep learning to address PPG’s temporal variability. CNN-LSTM architectures dominate the field:~\cite{8350983,8607019} achieved $\sim$96\% accuracy on wrist-worn PPG (TROIKA dataset, 125\,Hz) by combining spatial-temporal feature extraction. However, these models require high-frequency data and complex architectures (e.g., multi-layer CNNs), increasing computational costs. While~\cite{8553585} applied CNNs directly to raw PPG for real-time authentication (AUC: 78\%--83\%), their focus on accuracy overlooked energy efficiency—a critical gap for smartwatch and consumer-grade wearable devices requiring multi-day operation.

\subsection{Energy Efficiency and Low-Frequency PPG}
Efforts to optimize PPG systems for wearable devices have focused on adaptive sampling~\cite{zhao2021robust} and noise reduction~\cite{10109506}, but few address authentication. For example,~\cite{6004938} and~\cite{zhao2021robust} validated PPG’s usability in ambulatory settings but retained high sampling rates, limiting deployment in energy-sensitive scenarios like dementia care. Meanwhile, low-frequency PPG has been underexplored outside heart rate monitoring. While~\cite{4353358} tested derivatives of PPG for biometrics at 25\,Hz, their single-channel setup suffered from motion artifacts, and~\cite{9413906} improved robustness through 100\,Hz multichannel fusion but ignored energy costs. These gaps highlight the need for systems balancing fidelity, efficiency, and clinical utility.


\section{Conclusion}

In this paper, we presented a continuous user authentication system using only low-frequency (25\,Hz) multi-channel PPG signals from a wearable device. Despite the dramatically reduced sampling rate, our deep Bi-LSTM+Attention model achieved high authentication performance: about 88.11\% classification accuracy with low FAR of 0.48\%, FRR of 11.77\%, and EER of 2.76\% on a custom 26-user dataset. This accuracy is comparable to state-of-the-art PPG authentication approaches that rely on much higher sampling frequencies (75--500\,Hz), yet substantially reducing power consumption, yielding up to approximately 53\% savings in sensor energy use. The success of the model at only 25\,Hz---combined with training on an activity-diverse dataset-- demonstrated that our approach is robust against artifacts and varying real-world conditions. The long-term study on the We‑Be Band further confirmed that a single battery charge supports 25.8\,hours of continuous sensing and day‑long practicality in real‑world use. Furthermore, we implemented the entire system for real-time operation on the We-Be Band, and a deployment case study with 20 users confirmed the model’s effectiveness and stability in live use. To the best of our knowledge, this is the first work to fully implement and evaluate a continuous biometric authentication system on a smartwatch, highlighting the practicality and novelty of our on-device solution.

\balance
\section{Ethics Considerations}
This study involved the collection of biometric data—including photoplethysmography (PPG)—from human participants to develop and evaluate authentication mechanisms. Prior to data collection, we obtained approval from the Institutional Review Board (IRB). All participants provided informed consent after being briefed on the study's objectives, procedures, potential risks, and their rights, including the right to withdraw at any time without penalty.

To safeguard participant privacy, all collected data were anonymized by removing personally identifiable information (PII) and assigning randomized identifiers. Data storage and processing were conducted on secure, access-controlled systems in compliance with institutional and federal data protection guidelines. The study was designed to minimize risk, ensuring that procedures were non-invasive and posed no more than minimal risk to participants.

\bibliographystyle{IEEEtranS}
\input{main.bbl}

\end{document}